\documentclass[aps,twocolumn,amssymb,amsmath,superscriptaddress]{revtex4-1}

\usepackage{bm}
\usepackage{color}
\usepackage{graphicx}
\usepackage{dcolumn}
\usepackage{graphicx}
\usepackage[colorlinks=true, letterpaper=true, pdfstartview=FitV, linkcolor=blue, citecolor=blue, urlcolor=blue]{hyperref}
\usepackage[normalem]{ulem}
\usepackage{amsmath}
\usepackage{epstopdf}
\usepackage{multirow}
\usepackage{slashed}
\usepackage{float}

\setcitestyle{square}

\makeatletter
\renewcommand\@biblabel[1]{#1.}
\makeatother

\begin{document}

\title{Anomalous spatial shifts in interface electronic reflection beyond linear approximation}

\author{Runze Li}
\affiliation{Centre for Quantum Physics, Key Laboratory of Advanced Optoelectronic Quantum Architecture and Measurement (MOE), School of Physics, Beijing Institute of Technology, Beijing, 100081, China}
\affiliation{Beijing Key Lab of Nanophotonics \& Ultrafine Optoelectronic Systems, School of Physics, Beijing Institute of Technology, Beijing, 100081, China}

\author{Chaoxi Cui}
\affiliation{Centre for Quantum Physics, Key Laboratory of Advanced Optoelectronic Quantum Architecture and Measurement (MOE), School of Physics, Beijing Institute of Technology, Beijing, 100081, China}
\affiliation{Beijing Key Lab of Nanophotonics \& Ultrafine Optoelectronic Systems, School of Physics, Beijing Institute of Technology, Beijing, 100081, China}

\author{Xinxing Zhou}\email{xinxingzhou@hunnu.edu.cn}
\affiliation{Key Laboratory of Low-Dimensional Quantum Structures and Quantum Control of Ministry of Education, Synergetic Innovation Center for Quantum Effects and Applications, School of Physics and Electronics, Hunan Normal University, Changsha 410081, China}
\author{Zhi-Ming Yu}\email{zhiming\_yu@bit.edu.cn}
\affiliation{Centre for Quantum Physics, Key Laboratory of Advanced Optoelectronic Quantum Architecture and Measurement (MOE), School of Physics, Beijing Institute of Technology, Beijing, 100081, China}
\affiliation{Beijing Key Lab of Nanophotonics \& Ultrafine Optoelectronic Systems, School of Physics, Beijing Institute of Technology, Beijing, 100081, China}
%\author{Yugui Yao}
%\email{ygyao@bit.edu.cn}
%\affiliation{Centre for Quantum Physics, Key Laboratory of Advanced Optoelectronic Quantum Architecture and Measurement (MOE), School of Physics, Beijing Institute of Technology, Beijing, 100081, China}
%\affiliation{Beijing Key Lab of Nanophotonics \& Ultrafine Optoelectronic Systems, School of Physics, Beijing Institute of Technology, Beijing, 100081, China}

\begin{abstract}
Recently, the electronic analogy of the anomalous spatial shift, including Goos-H\"{a}nchen  and Imbert-Fedorov  effects, has been attracting widespread interest. The current  research  on the anomalous spatial shift in interface electronic reflection  is based on the paradigm of linear approximation, under which the center position of the incident and  reflected beams are obtained by  expanding the phases of relevant basis states and scattering amplitudes to the first order of  incident momentum.
However, in a  class  of normal cases, the linear  approximation can lead to  a divergent spatial shift in reflection for certain incident angles even though   the corresponding reflection possibility is finite.
In this work, we show  that such  non-physical results are caused by   an abrupt change in  the  number of the propagating  states at critical parameters, and can be resolved by calculating the center  positions of the  scattering beams beyond the linear approximation.
Moreover, we find that the beam width has an important influence on the spatial shift near the critical angles.
We demonstrate our idea via concrete calculations of Goos-H\"{a}nchen and Imbert-Fedorov shift on two representative models.
These results are beneficial for clarifying the scope of  application of the linear approximation in the study of anomalous spatial shifts.
%and for accurate description of the motion in waveguides.
\end{abstract}

\maketitle

\section{INTRODUCTION}
According to the laws of reflection in geometric optics, one knows that the  incident  point is always the same as the point where the  light beam is reflected back at a sharp interface.
However, due to the wave nature of photons,  these law should be revised in certain cases and  a light beam  can experience an anomalous spatial shift under reflection, namely,  there exists a shift between  the incident and the reflected beams at the  interface~\cite{Bliokh_2013}.
Generally, the spatial shift is  divided into longitudinal and transverse components with respect to the incident plane, known as   the  Goos-H\"{a}nchen shift~\cite{Goos1947,Renard1964} and Imbert-Fedorov shift~\cite{IF1955,IF1972,Onoda2004}, respectively. Since the wave-particle duality is a foundational concept in physics and holds for all particles, the anomalous spatial shift can also be found in many other particles, such as electrons~\cite{Miller1972,Beenakker2009,Chen2013,Wu2011,Yu2019}, atoms~\cite{JH2008} and neutrons~\cite{Haan2010}.

In electronic systems, the valence and conduction bands can  cross around Fermi level,  leading to non-trivial band degeneracy~\cite{Chiu2016,Armitage2018,Wan2011,yu2016,Weng2015,Wu2018,Soluyanov2015,Lix2021,2Lix2021}.
The band degeneracies in three-dimensional (3D) topological semimetals have many different types, and  can be classified as 0D nodal point, 1D nodal line and 2D nodal surface~\cite{yu2022,Guibin2022,Zeying2022}.
Remarkably, in the interface constructed by topological semimetals and other systems, both longitudinal and transverse shift effects are generally significant, due to strong (pseudo-)spin-orbit coupling in topological semimetals~\cite{Beenakker2009,Jiang2015,Yang2015,Yao2019,Feng2020}.
Such significant anomalous spatial shift can lead to various physical consequences, such as chirality-dependent Hall effect~\cite{Yang2015} and modifying  the dispersion of  the confined waveguide modes~\cite{Beenakker2009}.
Moreover, the behavior of the longitudinal and transverse shifts in these systems has a strong dependence on the species of the band degeneracies~\cite{Jiang2015,Yang2015,Yao2019,Feng2020}.
For example, when a beam comes from normal metal onto the interface with topological Weyl semimetals, there will exist quantum vortices in the vector field of the spatial shift in the interface momentum space, and the number of the quantum vortices is determined by the topological charge of the Weyl points~\cite{Yliu2020}.
The anomalous shifts can also be realized in Andreev reflection, during which  the incident particle changes its identity from electron to hole~\cite{Yliu2017,1Yliu2018,Yu2018,2Yliu2018}.
Similarly, the shifts strongly depend on the  pair potential of the superconductors, which in turn can be used to probe the superconducting states.

Currently, the  standard and the most general approach used to  study the anomalous shifts in electronic systems is the quantum scattering approach under linear approximation~\cite{Yu2019}.
In this approach, the incident beam  is modeled by wave packet $\Psi$, which is constructed by the incident  basis states $\psi^{i}(\boldsymbol{k})$ and is confined in both real and momentum spaces ($\boldsymbol{r}^{c}, \boldsymbol{k}^{c}$).
During scattering, the wave packet would be reconstructed, as each incident  basis state $\psi^{i}(\boldsymbol{k})$ is scattered into reflected basis state  $\psi^{r}(\boldsymbol{k})$ with certain  reflection amplitude.
The anomalous shifts then are obtained by comparing the center position of the incident and reflected beams.
In practice, one generally chooses the wave-packet profile to have a Gaussian form, and in such case, the anomalous shifts can be analytically obtained by expanding the phases  of  the relevant scattering basis states [$\psi^{i}(\boldsymbol{k})$ and $\psi^{r}(\boldsymbol{k})$] and reflection  amplitudes  to the linear order around $\boldsymbol{k}^{c}$.

The linear approximation is valid for most cases and gives accurate analytical results, which are helpful for gaining insight into  the  physics of the shifts.
However, for a class of normal cases, the linear approximation leads to a divergence of the anomalous  shifts in reflection at certain incident angles, even when the corresponding reflection possibility is finite~\cite{Beenakker2009,Jiang2015,Yu2018}. Then two important questions arise:  \emph{Under  what conditions, does the linear approximation not apply?  And how to  resolve the divergence of the shifts?}

In this work, we show that the divergence of the shifts in linear approximation are closely related to an abrupt change of the number of  propagating states.
In scattering, while the number of the scattering states is fixed, the number of the propagating states is not, and may abruptly changes when some critical  parameters like incident angle or Fermi energy change.
This abrupt change  would lead to a singularity in reflection amplitude or its derivative, which in the framework of linear approximation inevitably results in divergent anomalous  shifts.
We show this divergence can be resolved by calculating the centre position of the incident and reflected beams beyond the linear approximation.
We explicitly demonstrate our idea by calculating the longitudinal (Goos-H\"{a}nchen) and  transverse (Imbert-Fedorov) shifts on
two representative models.
Our work will be beneficial for clarifying the scope of application of the linear approximation in the study of anomalous spatial shifts.

\section{Quantum scattering approach} \label{SecII}
\begin{figure}[t]
\includegraphics[width=8.7cm]{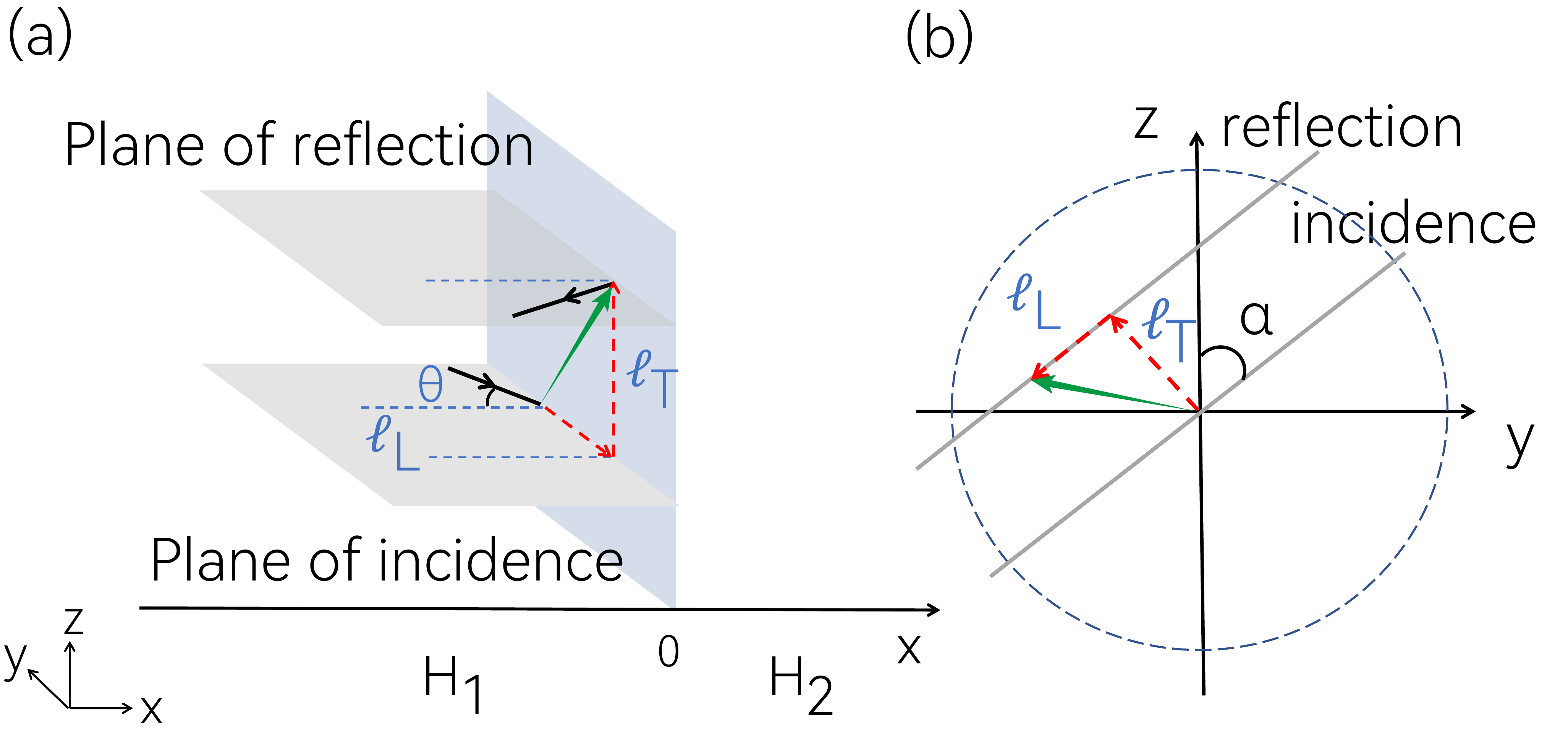}
\caption{(a) Schematic figure showing the anomalous spatial shift $\boldsymbol{\ell}$ in interface scattering. (b) Top view of the $y$-$z$ plane in (a).
\label{fig1}}
\end{figure}

Consider a general model which contains two media  respectively described  by Hamiltonian  $H_1$ and $H_2$, and a flat interface between these two media, as illustrated in Fig.~\ref{fig1}(a).
We also assume that the junction model is extended along $y$ and $z$ directions, indicating that  $k_y$ and $k_z$ are  conserved quantities during scattering.
A beam of particles is incoming  from the region of $x<0$ ($H_1$) and is scattered  at the interface residing at $x=0$ plane.
Besides, there is a rotation angle $\alpha$ between the incident plane and the $y$ axis, as shown in Fig.~\ref{fig1}(b).

To define the anomalous  spatial shift, the incident beam should be modeled by a wave packet, which is required to be confined in both real and momentum spaces.
%Without loss of generality, we chooses the wave-packet profile to have a Gaussian form, as the anomalous shift does not depend on the  shape of the wave packet.
We choose the wave-packet profile to have a Gaussian form.
Then, an incident wave packet centred at $\boldsymbol{k}^c=(k_y^c,k_z^c)$ can be written as \cite{Beenakker2009,Yu2019}
\begin{equation}\label{i beam}
\Psi^{i}(\boldsymbol{r},\boldsymbol{k}^c)=\int  w\left(\boldsymbol{k}-\boldsymbol{k}^c\right) \psi^{i}(\boldsymbol{k})  d \boldsymbol{k},
\end{equation}
where $\psi^{i}(\boldsymbol{k})=e^{i\boldsymbol{k} \cdot \boldsymbol{r}}|u^i(\boldsymbol{k})\rangle$ is the Bloch eigenstate of
the incident medium and $|u^i(\boldsymbol{k})\rangle$ is the cell-periodic part of the eigenstate.
The wave-packet profile $w$ reads
\begin{equation}
w(\boldsymbol{k})=\prod \limits_{i=y,z}\left(\sqrt{2 \pi} W_{i}\right)^{-1} e^{-k_{i}^{2} /\left(2 W_{i}^{2}\right)},
\end{equation}
where  $W_{i}$ denotes the Gaussian  width for the $i$-th component, controlling  the beam width in momentum space.
The centre position of the incident beam  can be written as~\cite{Luo2011,Ling2021}
\begin{eqnarray}
\boldsymbol{r}_{i}^{c}(\boldsymbol{k}^{c})=\frac{\int\boldsymbol{r}|\Psi^{i}(\boldsymbol{r},\boldsymbol{k}^{c})|^{2}d\boldsymbol{r}}{\int|\Psi^{i}(\boldsymbol{r},\boldsymbol{k}^{c})|^{2}d\boldsymbol{r}}.
\end{eqnarray}

When the incident beam hits the interface, each partial wave $\psi^{i}(\boldsymbol{k})$ is scattered into reflected basis state  $\psi^{r}(\boldsymbol{k})$ with certain $\boldsymbol{k}$-dependent  reflection amplitude $A_r(\boldsymbol{k})$.
The reflection amplitude here  is obtained by the standard  quantum scattering approach.
Then, the reflected beam  also is a wave packet and can be expressed as
\begin{equation}\label{i beam}
\Psi^{r}(\boldsymbol{r},\boldsymbol{k}^c)=\int  w\left(\boldsymbol{k}-\boldsymbol{k}^c\right) A_r(\boldsymbol{k}) \psi^{r}(\boldsymbol{k})  d \boldsymbol{k},
\end{equation}
and its centre position is
\begin{eqnarray}
\boldsymbol{r}_{r}^{c}(\boldsymbol{k}^{c})=\frac{\int\boldsymbol{r}|\Psi^{r}(\boldsymbol{r},\boldsymbol{k}^{c})|^{2}d\boldsymbol{r}}{\int|\Psi^{r}(\boldsymbol{r},\boldsymbol{k}^{c})|^{2}d\boldsymbol{r}}.
\end{eqnarray}
By comparing the centre position  of incident and reflected beam, the anomalous shift in reflection  is obtained as
\begin{eqnarray} \label{exact_1}
\boldsymbol{\ell}(\boldsymbol{k}^{c})=\boldsymbol{r}_{r}^{c}(\boldsymbol{k}^{c})-\boldsymbol{r}_{i}^{c}(\boldsymbol{k}^{c}),
\end{eqnarray}
which is a vector and forms a vector field in the interface momentum space.
However, it should be noticed that based on Eq.~(\ref{exact_1}), while one always can numerically obtain  $\boldsymbol{\ell}(\boldsymbol{k}^{c})$,  it is impossible to obtain an analytical expression for $\boldsymbol{\ell}(\boldsymbol{k}^{c})$, which prevents a deep understanding of the physics underlying the shifts.
To resolve this problem, one  has to  resort to  linear approximation, as it can give an  analytical expression of  $\boldsymbol{\ell}(\boldsymbol{k}^{c})$ in many cases.

The Bloch eigenstate of the incident  medium  may include multiple ($N$) components, i.e. $|u^{i(r)}(\boldsymbol{k})\rangle=(u^{i(r)}_1,\cdots u^i_N)^T$, then the   wave packets $\Psi^{i(r)}$  should also have  $N$ components.
The $n$-th component of $\Psi^{i}$ and $\Psi^{r}$  are respectively  expressed as
\begin{eqnarray}\label{int_Psi}
\Psi^{i}_{n}(\boldsymbol{r},\boldsymbol{k}^c)&=&\int  w\left(\boldsymbol{k}-\boldsymbol{k}^c\right) e^{i\boldsymbol{k} \cdot \boldsymbol{r}} |u^i_n(\boldsymbol{k})| e^{i \phi^{i}_n(\boldsymbol{k})}  d \boldsymbol{k},
\end{eqnarray}
 and
\begin{eqnarray}\label{ref_Psi}
\Psi^{r}_{n}(\boldsymbol{r},\boldsymbol{k}^c)&=&\int  w\left(\boldsymbol{k}-\boldsymbol{k}^c\right) |A_r| e^{i \varphi(\boldsymbol{k})} e^{i\boldsymbol{k} \cdot \boldsymbol{r}} |u^r_n(\boldsymbol{k})| e^{i \phi^{r}_n(\boldsymbol{k})}  d \boldsymbol{k}, \nonumber \\
\end{eqnarray}
with $\phi^{i(r)}_n(\boldsymbol{k})=\arg [u^{i(r)}_n(\boldsymbol{k})]$ and  $\varphi(\boldsymbol{k})=\arg [A_r(\boldsymbol{k})]$.
In the framework of  linear approximation, the two phases  $\phi^{i(r)}_n(\boldsymbol{k})$ and $\varphi(\boldsymbol{k})$ in Eq. (\ref{int_Psi}) and Eq. (\ref{ref_Psi})   are replaced by the first order Taylor series expanded around $\boldsymbol{k}^c$,
\begin{eqnarray}\label{phase1}
\phi^{i(r)}_n(\boldsymbol{k})&=&\phi^{i(r)}_n(\boldsymbol{k}^c)+(\boldsymbol{k}-\boldsymbol{k}^c)\cdot \partial_{\boldsymbol{k}}\phi^{i(r)}_n|_{\boldsymbol{k}=\boldsymbol{k}^c},  \\
\varphi(\boldsymbol{k})&=&\varphi(\boldsymbol{k}^c)+(\boldsymbol{k}-\boldsymbol{k}^c)\cdot \partial_{\boldsymbol{k}}\varphi|_{\boldsymbol{k}=\boldsymbol{k}^c}.
\end{eqnarray}
With the linear approximation, one can find that Eqs.~(\ref{int_Psi}) and~(\ref{ref_Psi}) will take the following forms
\begin{eqnarray}
\Psi^{i}_{n}(\boldsymbol{r},\boldsymbol{k}^c)& \propto& e^{-\sigma_y^2 (y+\partial_{k_y}\phi^{i}_n|_{\boldsymbol{k}=\boldsymbol{k}^c})^2 /2} \nonumber \\
&&\times e^{-\sigma_z^2 (z+\partial_{k_z}\phi^{i}_n|_{\boldsymbol{k}=\boldsymbol{k}^c})^2/2},\\
\Psi^{r}_{n}(\boldsymbol{r},\boldsymbol{k}^c)& \propto& e^{-\sigma_y^2 (y+\partial_{k_y}(\phi^{r}_n+\varphi)|_{\boldsymbol{k}=\boldsymbol{k}^c})^2 /2} \nonumber \\
&&\times e^{-\sigma_z^2 (z+\partial_{k_z}(\phi^{r}_n+\varphi)|_{\boldsymbol{k}=\boldsymbol{k}^c})^2 /2},
\end{eqnarray}
indicating that $\Psi^{i}_{n}$ and $\Psi^{r}_{n}$ are centred at $-\partial_{\boldsymbol{k}}\phi^{i}_n|_{\boldsymbol{k}=\boldsymbol{k}^c}$  and $-\partial_{\boldsymbol{k}}(\phi^{r}_n+\varphi)|_{\boldsymbol{k}=\boldsymbol{k}^c}$, respectively.
The centre position of  incident and reflected beams are the average of  all components, written as
\begin{eqnarray}
\boldsymbol{r}^{c}_i& =& -\sum_{n} w_n^{i} \partial_{\boldsymbol{k}}\phi^{i}_n|_{\boldsymbol{k}=\boldsymbol{k}^c}\\
\boldsymbol{r}^{c}_r& =& -\sum_{n} w_n^{r} \partial_{\boldsymbol{k}}(\phi^{r}_n+\varphi)|_{\boldsymbol{k}=\boldsymbol{k}^c},
\end{eqnarray}
with $w_n^{i(r)}$ the weight  of the  $n$-th component of $|u^{i(r)}\rangle$, satisfying $\sum_n (w_n^{i(r)})^2=1$.
Hence, the anomalous shift from linear approximation  can be analytically expressed as the difference between the two centre positions,
\begin{eqnarray} \label{ellLA}
\boldsymbol{\ell}^{LA}& =& \sum_{n} [w_n^{i} \partial_{\boldsymbol{k}}\phi^{i}_n|_{\boldsymbol{k}=\boldsymbol{k}^c}-w_n^{r} \partial_{\boldsymbol{k}}(\phi^{r}_n+\varphi)|_{\boldsymbol{k}=\boldsymbol{k}^c}].
\end{eqnarray}
By analyzing  the above derivation, we find that there two prerequisites for application of the linear approximation. (1) After  scattering, the reflected beams should be  still centred at $\boldsymbol{k}^{c}$ in momentum space. (2) The reflection amplitude $A_r(\boldsymbol{k})$ should be an  analytic function in the neighbourhood of $\boldsymbol{k}^{c}$, as it is a  prerequisite for the Taylor expansion of $\arg [A_r(\boldsymbol{k})]$.

These two conditions can be satisfied for most cases.
Generally, condition (1) is always satisfied except that the reflection probability  is vanishing at $\boldsymbol{k}=\boldsymbol{k}^{c}$.
But, in such case, the anomalous shift would  be irrelevant for  physical observations, as no particle is reflected back and then the anomalous shift will not happen.
In contrast,  we  show  that  the condition (2) does  not hold at certain  incident angles and model parameters, beyond which the number of the scattered propagating states changes.
This abrupt change generally leads to divergent shifts [obtained from linear approximation (\ref{ellLA})] at the critical parameters, which apparently are not correct.
Hence, to obtain correct results, we have to calculate the anomalous  shifts via  Eq.~ (\ref{exact_1}). We find that at the critical parameters, the anomalous shifts obtained from Eq.~(\ref{exact_1}) are completely different from that obtained from linear approximation ~(\ref{ellLA}), and are  not divergent. Away from the critical parameters, the  anomalous shifts obtained from both approaches are almost the same.
Besides, the anomalous shifts around (away from) the critical parameters  have a strong (negligible) dependence on the beam width $W$. % and the  shape of the wave packet.
 In the following, we use two representative examples to demonstrate our ideas.

\section{Longitudinal shift in graphene model} \label{SecIII}
In the first case, we consider a graphene junction model. This model is simple and features a divergent longitudinal spatial  shift with finite reflection probability~\cite{Beenakker2009}.
However, the origin of the divergence has not been discussed. Here, we will show the divergence is caused by an  abrupt change of scattering environment.

According to the setup in Fig.~\ref{fig1}, we assume the junction model is lied on the $x$-$y$ plane. Notice that  the $z$-direction here is a dummy degree of freedom, as  graphene is a 2D system and the longitudinal shift is within the scattering plane. The Hamiltonian of the junction model is  given as
\begin{equation}\label{Model1}
\mathcal{H}= \begin{cases} v k_x\sigma_x  +v k_y\sigma_y, & x < 0, \\ v k_x\sigma_x  +v k_y\sigma_y+V, & x>0,\end{cases}
\end{equation}
where $v$ is the fermi velocity, $\mathbf{\sigma}$ is the Pauli matrix and $V$ denotes a potential energy applied on the $x>0$ region.

\begin{figure}[t]
\includegraphics[width=8.7cm]{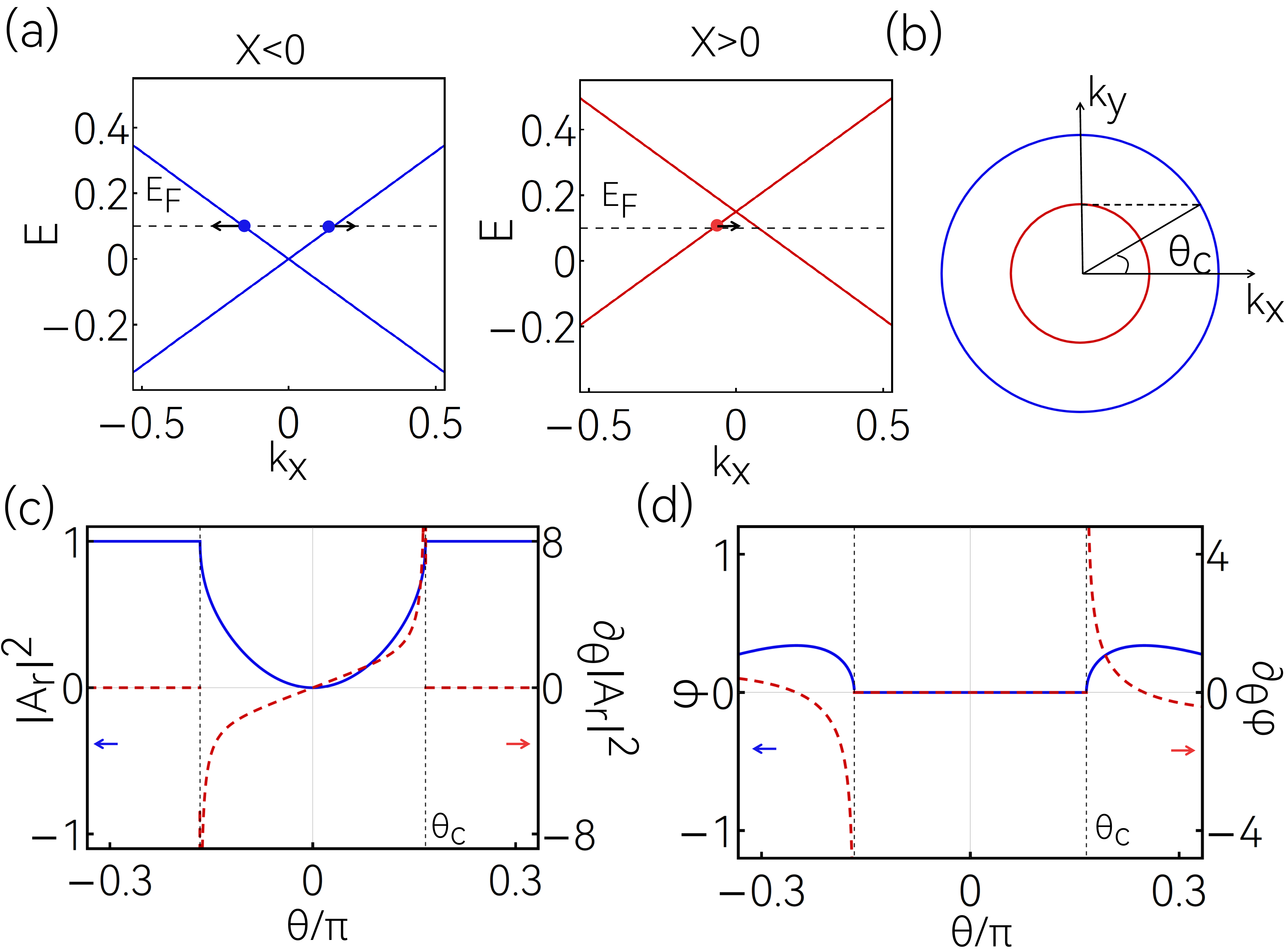}
\caption{(a) The band structure of incident medium ($x<0$) and transmitted medium  ($x>0$) in
model (\ref{Model1}).
The spheres denotes the incident, reflected and transmitted  electron  state, and the arrows indicate their
 moving directions.
(b) The Fermi surfaces of both $x<0$ (blue circle) and  $x>0$ (red circle)  medium with the label of the  critical   angel $\theta_c$.
(c) $|A_r|^2$ and its derivative  and (d)  the phase $\varphi=\arg (A_r)$ and its derivative vs incident angle $\theta$. In the calculation,
we choose $E_F=100\ \rm{meV}$, $V=150\ \rm{meV}$ and $v=10^6\ \rm{m/s}$. \label{fig2}}
\end{figure}

We  plot the band structure of the graphene model  in both $x<0$ and $x>0$ regions in Fig.~\ref{fig2}(a) and the corresponding equienergy contours in Fig.~\ref{fig2}(b).
The possible incident  and reflected electron states are also marked in Fig. \ref{fig2}(a).
One can find that for a fixed Fermi energy $E_F>0$  (measured from the Dirac point) and a finite $V$ satisfying  $0<V<2E_F$, there exists two critical incident angle $\pm \theta_c$, beyond which there is no propagating mode  for transmitted state [see Fig.~\ref{fig2}(b)].
The critical angle is given as
\begin{equation}
\theta_c=\arcsin \left|\frac{V-E_F}{E_F}\right|.
\end{equation}
When the incident angle $\theta=\arctan (k_y/k_x)$ is smaller than  the critical angle $|\theta|<\theta_c$, an incident electron from $x<0$ region can be reflected  (transmitted)  as a propagating state in $x<0$ ($x>0$) region.
However, the number of transmitted propagating states varies from one to zero when $|\theta|>\theta_c$, indicating an  abrupt change of scattering environment.
As discussed in Ref.~{\cite{Ma2018}}, this change generally results in a discontinuity in the derivative of the reflection amplitude.
As a consequence, the Taylor expansion of the phase of  the reflection amplitude around the critical incident angles $\pm\theta_c$  will be meaningless, making   the  anomalous spatial shifts obtained by  linear approximation inaccurate.

To directly show this, we proceed to solve the scattering states of the graphene junction model (\ref{Model1}), which can be written as
\begin{equation}
\psi(\boldsymbol{k})= \begin{cases} \psi^{i}(\boldsymbol{k})+A_r \psi^{r}(\boldsymbol{k}), & x < 0, \\ A_t \psi^{t}(\boldsymbol{k}), & x>0,\end{cases}
\end{equation}
where $A_{r(t)}$ is the reflection (transmission) amplitude, and $\psi^{i}$, $\psi^{r}$ and $\psi^{t}$ are the basis state for incident, reflected and transmitted states, respectively.
Explicitly, the basis states read
\begin{eqnarray}
\psi^{i}(\boldsymbol{k})&=&\frac{1}{\sqrt{2}}\left(\begin{array}{c}
e^{-i \theta / 2} \\
 e^{i \theta / 2}
\end{array}\right) e^{i k_{i} x+i k_{y} y}, \\
\psi^{r}(\boldsymbol{k})&=&\frac{1}{\sqrt{2}}\left(\begin{array}{c}
-ie^{i \theta / 2} \\
i e^{-i \theta / 2}
\end{array}\right) e^{-i k_{i} x+i k_{y} y},\\
\psi^{t}(\boldsymbol{k})&=&\frac{1}{\sqrt{2}|E_F-V|}\left(\begin{array}{c}
E_F-V \\
 v (k_t+ i k_y)
\end{array}\right) e^{i k_{t} x+i k_{y} y}.
\end{eqnarray}
Here $k_i=v^{-1}\sqrt{E_F-v^2 k_y^2}$,  $\theta=\arctan{(k_y/k_i)}$ is the incident angle,   $k_t=v^{-1}\text{sgn}(E_F-V)\sqrt{(E_F-V)^2-v^2 k_y^2}$ for $|\theta|<\theta_c$ and $k_t=i\kappa$ for $|\theta|>\theta_c$, where $\kappa=v^{-1}\sqrt{v^2 k_y^2-(E_F-V)^2}$.
For the graphene junction model here, the  boundary condition at the interface is
\begin{equation}
\psi(x=0^-)=\psi(x=0^+),
\end{equation}
with which  the reflection amplitude $A_r$ is obtained as
\begin{equation}
A_r=\frac{v(\kappa+k_y)+ i e^{i\theta }(E_F-V)}{v e^{i\theta }(i \kappa+i k_y)+(E_F-V)},
\end{equation}

The square of the module of $A_r$ ($|A_r|^2$) and the phase $\varphi=\arg (A_r)$ as functions of the incident angle are plotted in Fig.~\ref{fig2}(c) and Fig.~\ref{fig2}(d), respectively. One observes that for  $|\theta|<\theta_c$, $A_r$ is a finite real number with $\varphi=0$, and for  $|\theta|>\theta_c$, $A_r$ becomes a complex number with $|A_r|^2=1$, indicating the appearance of total reflection.
This is  consistent with the fact that when $|\theta|>\theta_c$, there no longer  exists a  propagating mode for transmitted state.
Importantly, while both $|A_r|^2$ and $\varphi$ are continuous functions, their derivative  are discontinuous at the critical angle $\pm\theta_c$ [see Fig.~\ref{fig2}(c) and (d)], corresponding to an abrupt change of the scattering condition, namely, the disappearance of the transmitted propagating mode for $|\theta|>\theta_c$.

Based on the expression of $A_r$,  the longitudinal shift under linear approximation is  obtained as
\begin{equation} \label{LAy_G}
\ell_y^{LA}=2\frac{\sin ^2 \theta+1-V / E_F}{\kappa \sin 2\theta }.
\end{equation}
As discussed above, the accuracy of the anomalous shift obtained from linear approximation requires $A_r$ to be an analytic function. But $A_r$ is not an analytic function in the neighbourhood of $\theta=\pm\theta_c$. Hence, Eq. (\ref{LAy_G}) may be inaccurate around $\pm \theta_c$.
\begin{figure}[t]
\includegraphics[width=8.7cm]{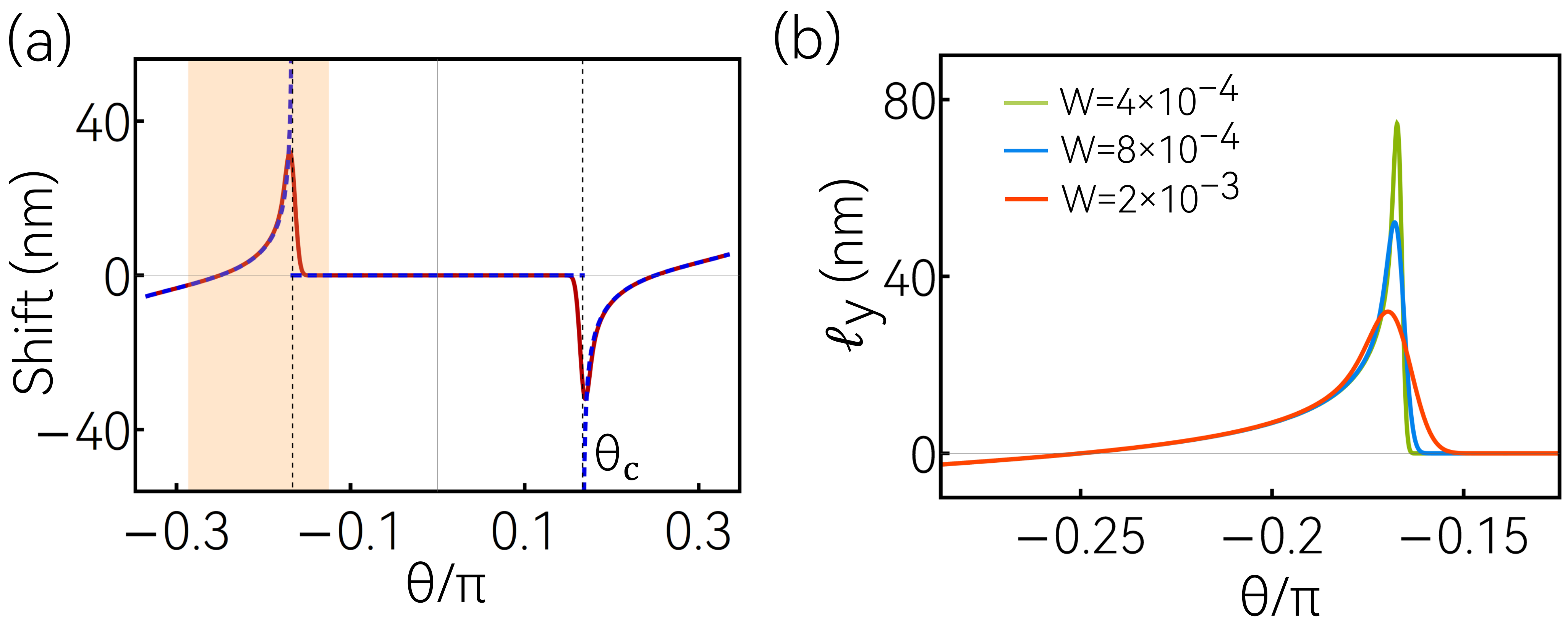}
\caption{The anomalous longitudinal shift in model (\ref{Model1}).
(a) The results obtained with linear approximation $\ell_y^{LA}$ (dashed line)
and that obtained without approximation  $\ell_y$ (solid line) vs incident angle.
(b) $\ell_y$ vs $\theta$ for different beam width $W$.
The box in (a) denotes the range of $\theta$ in (b). In the calculation,
we choose $E_F=100\ \rm{meV}$, $V=150\ \rm{meV}$, $v=10^6\ \rm{m/s}$.
\label{fig3}}
\end{figure}
In Fig.~\ref{fig3}(a), we plot the typical behavior of the longitudinal shift $\ell_y^{LA}$ from Eq.~(\ref{LAy_G}), along with the shift  $\ell_y$ obtained from Eq.~(\ref{exact_1}) beyond linear approximation.
We have checked that the wave packet of the reflected beam still features a Gaussian-type profile and is centred  at same momentum with the incident  wave packet.
Hence, the shift obtained from Eq.~(\ref{exact_1}) should be accurate.
From Fig. \ref{fig3}(a), one observes that the $\ell_y^{LA}$ has two discontinuity points at $\pm \theta_c$. Specifically, it is zero for  $|\theta|<\theta_c$,  diverges abruptly at $\pm \theta_c$ and  becomes  finite when  moving  away from  $\pm \theta_c$.
In contrast, $\ell_y$ obtained from Eq.~(\ref{exact_1}) is always continuous for any incident angle.
Particularly, it  is not divergent at $\pm \theta_c$, indicating the non-physical divergent results from linear approximation is resolved by using  Eq.~(\ref{exact_1})
to calculate the anomalous shift. Besides, there are several  other  observations.
(i) At critical angle $\pm \theta_c$, while $\ell_y$ is not divergent, it still  is very large and can reach few tens of   nanometer with reasonable  parameters.
(ii)  Away from $\pm \theta_c$, $\ell_y^{LA}$ and $\ell_y$ share similar behavior and are almost identical, as shown in Fig.~\ref{fig3}(a).

The anomalous shifts~(\ref{LAy_G}) obtained from linear approximation are  always independent of  the beam width ($\sim W=|\bm{W}|$). However, in optics, it has been  demonstrated  that the  anomalous shifts around  the critical parameters  strongly depend on the beam width \cite{Lai1986,Wang2013,Qiu2015}.
Here, we also calculate the longitudinal shift $\ell_y$ from Eq.~(\ref{exact_1}) for different $W$. The obtained results are  shown in Fig.~\ref{fig3}(b), from which one can find that the maximum value of $\ell_y$ indeed varies with  $W_y$, consistent with the results  in optics.
Again, when moving away from $\pm \theta_c$, $\ell_y$ is not sensitive to the value of $W$.

\section{Transverse shift in Andreev  reflection}
In the second case, we consider transverse shift in Andreev  reflection~\cite{Yliu2017,Yu2018}.
In electronic systems, besides ordinary electron scattering, Andreev  reflection is another intriguing scattering process that occurs at the interface between metal and superconductor and is described by the Bogoliubov-de Gennes (BdG) equation.
The junction model here is of a 3D electron gas interfaced with a $d$-wave superconductor. The corresponding BdG equation can be written as
\begin{equation}\label{AFHam}
H(x)\psi=\varepsilon \psi,
\end{equation}
with $\varepsilon$ the excitation energy
\begin{equation}
H=\left[\begin{array}{cc}
-\frac{1}{2 m} \nabla^{2}-E_{F} & 0 \\
0 & -\mathcal{T} (-\frac{1}{2 m} \nabla^{2}) \mathcal{T}^{-1}+E_{F}
\end{array}\right],
\end{equation}
for normal metal  region ($x<0$), and
\begin{equation}
H=\left[\begin{array}{cc}
H_{0}+V-E_{F} & \Delta(k_{\|}) \\
\Delta^{*}(k_{\|}) & -\mathcal{T} H_{0} \mathcal{T}^{-1}-V+E_{F}
\end{array}\right],
\end{equation}
for the superconductor region ($x>0$). Besides, the interface barrier potential $h \delta(x)$ is considered.
Here, $m$ denotes the electron mass, $E_F$ is the  Fermi energy, $\mathcal{T}$ is the time reversal operator, $H_{0}=-\frac{1}{2 m} \left(\partial_{z}^{2}+\partial_{y}^{2}\right)-\frac{1}{2 m_{x}} \partial_{x}^{2}$,  $\Delta(k_y,k_z)=\Delta_0 \cos(2 \phi_k)$ represents a $d_{y^2-z^2}$ pairing with  $\phi_k=\arctan(k_y/k_z)$, and $V$ is a potential energy ensuring  $E_F-V\gg \Delta_0$.
According to the setup  in  Fig.~\ref{fig1}, the rotation angle is identical to $\phi_k$, namely, $\alpha=\phi_k$.

This model has been used to show that transverse shift can be solely induced by the unconventional pairing, and it is found that by  varying  rotation angle $\alpha$ [see Fig.~ \ref{fig1}(b)], the transverse shift becomes  divergent at certain critical angles.
Here, we will show that the divergence is also caused by  the change in the number of transmitted propagating states at the critical angles.

\begin{figure}[t]
\includegraphics[width=8.7cm]{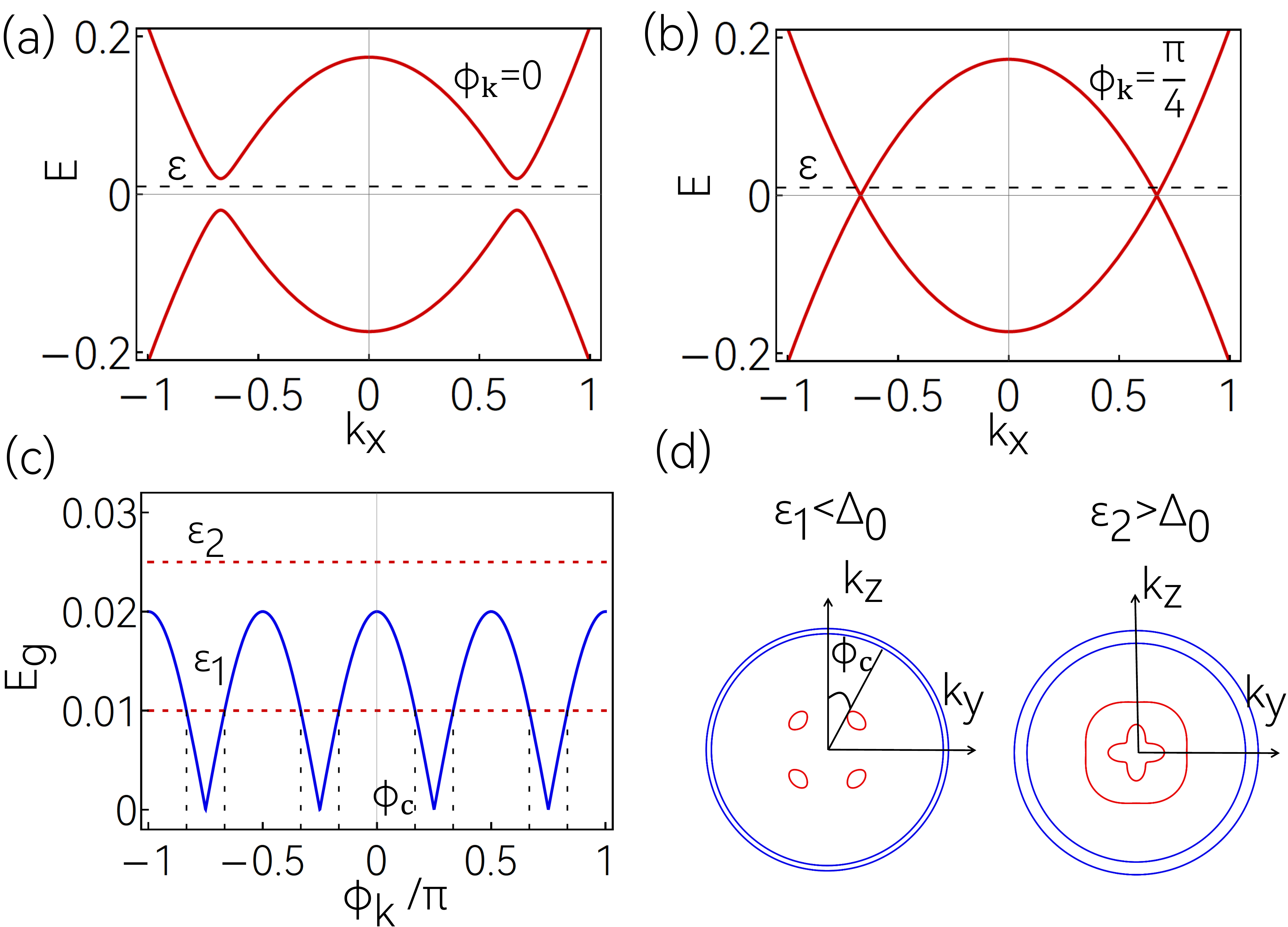}
\caption{Band structure of the BdG Hamiltonian in model (\ref{AFHam}).
(a), (b) show the BdG spectrum in different incident planes.
(c) denotes the band gap  $E_g$ of the  BdG spectrum.
(d) The BdG Fermi surfaces with different excitation energy. In the calculation,
we choose $\varepsilon=10\ \rm{meV}$, $\Delta_0=20\ \rm{meV}$, $E_F=0.4\ \rm{eV}$, $V=0.2\ \rm{eV}$, $h=0.3\ \rm{eV}\cdot\rm{nm}$ and the incident angle $\theta=\pi/12$. We take the $\varepsilon_1=10\ \rm{meV}$, $\varepsilon_2=25\ \rm{meV}$ in (c), (d).
\label{fig4}}
\end{figure}

The scattering states of Eq.~(\ref{AFHam})  can be written as~\cite{Yu2018}
\begin{equation}
\psi(\boldsymbol{k})= \begin{cases} \psi^{i}(\boldsymbol{k})+r_e \psi^{r}_e(\boldsymbol{k})+r_h \psi^{r}_h(\boldsymbol{k}), & x < 0, \\ t_1 \psi_1^{t}(\boldsymbol{k})+t_2 \psi_2^{t}(\boldsymbol{k}), & x>0,\end{cases}
\end{equation}
where $r_{e(h)}$ is the amplitude for the normal (Andreev) reflection, and $t_{1(2)}$ is the transmission amplitude, and the $\psi$ are the corresponding basis states, expressed as
\begin{eqnarray}
\psi^{r}_e(\boldsymbol{k})&=&\left(\begin{array}{l}
1 \\
0
\end{array}\right)e^{-i k_x^e x+i k_y y+i k_z z}, \\
\psi^{r}_h(\boldsymbol{k})&=& \left(\begin{array}{l}
0 \\
1
\end{array}\right)e^{i k_x^h x+i k_y y+i k_z z}, \\
\psi_{\pm}^{t}(\boldsymbol{k})&=& \left(\begin{array}{c}
1 \\
\eta_{\pm}
\end{array}\right)e^{i k_S^{\pm} x+i k_y y+i k_z z},
\end{eqnarray}
where $\eta_{\pm}=\frac{\Delta_{\boldsymbol{k}_{\pm}}}{\varepsilon\pm\sqrt{\varepsilon^{2}-\Delta_{k_{S}^{\pm}}^{2}}}$ with $\Delta_{\boldsymbol{k}_{\pm}}=\boldsymbol{\Delta}\left(k_{\pm}^{\mathrm{S}}, k_y, k_z\right)$,  $k_\|=\sqrt{k_y^2+k_z^2}$,  $k_x^{e / h}=\sqrt{2 m E_{F}-k^2_\|}$, and  $k_{S}^{\pm}=\pm\sqrt{2m_x(E_f-U- k^2_\|/2 m)}$.

Before processing to the concrete calculations of the scattering amplitudes and the anomalous shifts, we discuss the influences of rotation angle and the value of excitation energy on the scattering in the interface.
Since the $d$-wave superconductor is anisotropic in $k_y$-$k_z$ plane, the band structure of the superconductor in different  incident planes (determined by the rotation angle $\alpha$) will be different, as shown in Fig.~\ref{fig4}(a) and~\ref{fig4}(b), indicating the transverse shift is sensitive to the rotation angle $\alpha$ and $\phi_k$.
We also plot the band gap $E_{\text{gap}}=|\Delta_{\boldsymbol{k}}|$ of the superconductor as a function of $\phi_k$ in Fig.~\ref{fig4}(c), showing the superconductor becomes gapless at $\phi_k=\pm \pi/4, \ \pm 3\pi/4$.
Then, a key observation is that for any excitation  energy satisfying $|\varepsilon|<\Delta_0$, the two transmitted states ($\psi_{1}^{t}$ and $\psi_{2}^{t}$) are propagating modes for $|\cos 2\phi|<\varepsilon/\Delta_0$  ($|\varepsilon|>E_{\text{gap}}$) and are evanescent modes for $|\cos 2\phi|>\varepsilon/\Delta_0$ ($|\varepsilon|<E_{\text{gap}}$), as illustrated  in Fig. \ref{fig4}(c).
In contrast, the two transmitted states are always  propagating modes when $|\varepsilon|>\Delta_0$, as $|\varepsilon|>E_{\text{gap}}$ for any $\phi_k$ [see Fig. \ref{fig4}(d)].
Similarly, one can expect that the critical rotation angles for divergent transverse  shifts in Ref. {\cite{Yu2018}.} satisfy  $|\cos 2\phi_c|=\varepsilon/\Delta_0$ with  $|\varepsilon|<\Delta_0$.
Besides, when $|\varepsilon|>\Delta_0$, the transverse  shifts  obtained from linear approximation will be accurate and consistent with that obtained from Eq.~(\ref{exact_1}).

With the boundary conditions,
\begin{eqnarray}
\psi(x=0^-)&=&\psi(x=0^+),\\
\frac{1}{m}\partial_x\psi (x=0^-)&=&\frac{1}{m_x}\partial_x\psi (x=0^+)-h\psi (0),
\end{eqnarray}
the Andreev reflection amplitude $r_h$ is obtained as
\begin{equation}
r_h=\frac{-4 \eta_{-} \eta_{+} \Gamma}{\eta_{+} \left(Z-2 \Gamma\right)-\eta_{-} \left(Z+2 \Gamma\right)},
\end{equation}
with $Z=4 \frac{mh}{k_{e}}^{2}+1+\Gamma^{2}$ and $\Gamma=\frac{mk_{s}^{+}}{m_{1}k_{e}}$.

\begin{figure}[t]
\includegraphics[width=8.7cm]{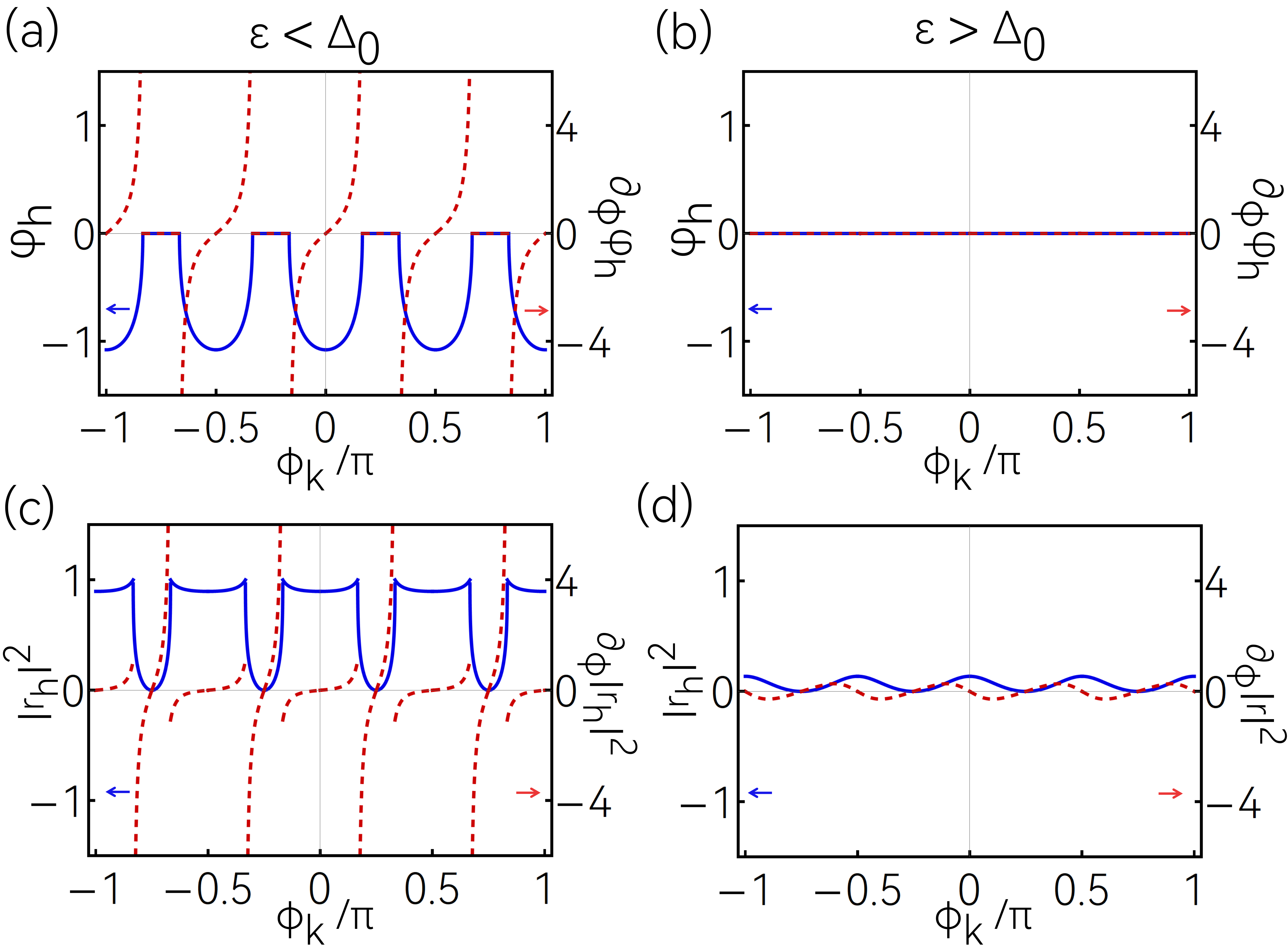}
\caption{
(a,b) The phase $\varphi_h=\arg (r_h)$ and its derivative and (c,d) $|r_h|^2$
and its derivative vs  $\phi_k$   in the cases of $|\varepsilon|<\Delta_0$ and
$|\varepsilon|>\Delta_0$. Here, we choose $\Delta_0=20\ \rm{meV}$, $E_F=0.4\ \rm{eV}$, $V=0.2\ \rm{eV}$, $h=0.3\ \rm{eV}\cdot\rm{nm}$ and $\theta=\pi/12$.  We take the $\varepsilon=10\ \rm{meV}$ in (a), (c) and $\varepsilon=30\ \rm{meV}$ in (b), (d). \label{fig5}}
\end{figure}

In Fig.~\ref{fig5}, we plot the  obtained  $|r_h|^2$ and  $\varphi_h=\arg (r_h)$ as functions of the rotation angle $\phi_k$  for $|\varepsilon|<\Delta_0$ and $|\varepsilon|>\Delta_0$.
We find that the derivative of $|r_h|^2$ and  $\varphi_h=\arg (r_h)$ exhibit eight discontinuity points at  $\phi_c=\pm [\arccos (\varepsilon/\Delta_0)]/2$ when $|\varepsilon|<\Delta_0$, but are smooth functions when $|\varepsilon|>\Delta_0$, consistent with the analysis of the  band structure of the junction model.
Interestingly, $r_h$ is a pure real number when $|\varepsilon|>\Delta_0$.
We then study the dependence of anomalous transverse shift on the rotation angle and $\phi_k$.

\begin{figure}[t]
\includegraphics[width=8.7cm]{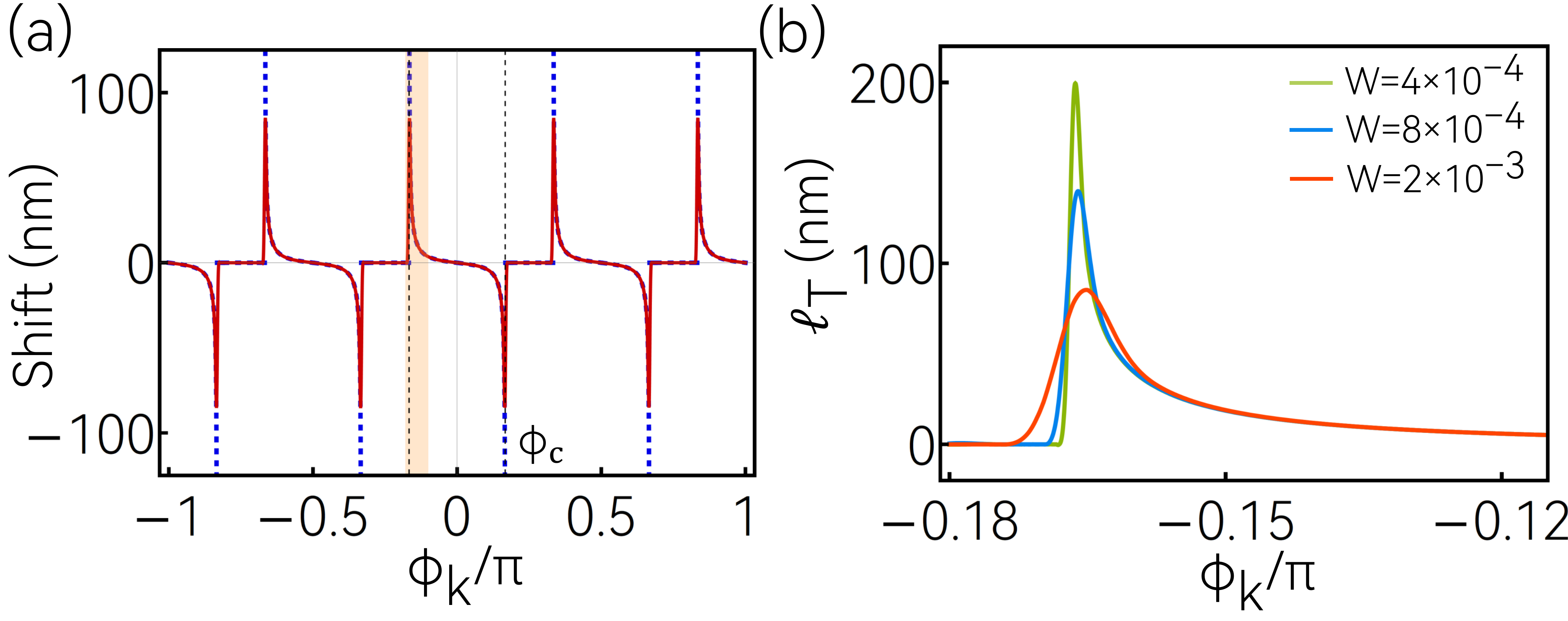}
\caption{
The anomalous transverse shift in model (\ref{AFHam}).
(a) The results obtained with linear approximation $\ell_T^{LA}$ (dashed line)
and that obtained without approximation  $\ell_T$ (solid line) vs $\phi_k$.
(b) $\ell_T$ vs $\phi_k$ for different beam width $W$.
The box in (a) denotes the range of $\phi_k$ in (b). Here, we choose $\varepsilon=10\ \rm{meV}$, $\Delta_0=20\ \rm{meV}$, $E_F=0.4\ \rm{eV}$, $V=0.2\ \rm{eV}$, $h=0.3\ \rm{eV}\cdot\rm{nm}$ and $\theta=\pi/12$.  \label{fig6}}
\end{figure}

Under the linear approximation, the anomalous transverse shift in Andreev reflection is established as
\begin{eqnarray}\label{TLA}
\ell_T^{LA}&=&-\frac{1}{k_\parallel}\partial_{\phi_k}\varphi_h \nonumber\\
&=&-\frac{\rho \Lambda \Delta_0^2 \sin (4 \phi_k)}{k_{\|} \varepsilon^2\left(1+\Lambda^2 / \rho^2\right)}  \Theta\left(\left|\Delta_0 \cos 2 \phi_k\right|-\varepsilon\right)
\end{eqnarray}
with $\rho=\varepsilon/\sqrt{\Delta_0^2 \cos ^2(2 \phi_k)-\varepsilon^2}$ and $\Lambda=Z/(2 \Gamma)$.
According to Eq. (\ref{TLA}), we find the transverse shift obtained from linear approximation $\ell_T^{LA}$ indeed is divergent at $\phi_c$ when $|\varepsilon|<\Delta_0$, as shown Fig. \ref{fig6}.
For $\varepsilon>\Delta_0$, the transverse shift will be zero, as  $r_h$ is a real number with $\varphi_h=0$.
%For comparison, we also plot  $\ell_T^{LA}$ as a function of  $\alpha$ for $|\varepsilon|>\Delta_0$, and find $\ell_T$  is not divergent for any $\alpha$.

The numerical results of the  transverse shifts $\ell_T$ obtained from Eq.~(\ref{exact_1}) with  $|\varepsilon|<\Delta_0$ and different beam width also are plotted in Fig.~\ref{fig6}.
We have checked that the wave packets of the reflected hole beam still  can be well defined and exhibit a Gaussian-type profile, indicating  the calculated anomalous  shift would be  reliable.
One observes that for all the beam widths, the  transverse shifts are not divergent at the critical angles $\phi_c$.
Similar to the first case, we find that (i) the anomalous shift indeed is significant at the critical angles and can reach few tens of nanometer with reasonable  parameters, (ii) $\ell_T$ and $\ell_T^{LA}$ are almost identical away from the critical angle $\phi_c$ and (iii) the anomalous shift is sensitive to the beam width when and only when $\phi_k$ is  close to  $\alpha_c$.
Besides, when $|\varepsilon|/\Delta_0>1$, the transverse shifts $\ell_T$ obtained from Eq.  ~(\ref{exact_1}) is negligible, consistent with the results from linear approximation.

\section{Conclusions}

In this work, we study the   anomalous shift in  interface electronic reflection based on the quantum scattered approach with and without linear approximation.
We find that for a large case of junction models, the propagating modes of scattering states may be changed by  varying  certain parameters like incident angle.
Around  the critical parameters, the linear approximation is invalid and  leads to divergent anomalous shifts in scattering.
In contrast, the    quantum scattered approach without linear approximation always gives accurate results,
which are  significant  but not divergent  around the  critical parameters.
Moreover, we show the anomalous shifts around the critical parameters  decrease when increasing the width of the incident beam.
This means that the narrower the incident beam, the more pronounced the anomalous shift.
Away from the critical parameters, the anomalous shifts obtained from the quantum scattered approach with and without linear approximation are similar and not sensitive to the beam width.

%At last, we would like to discuss a  possible application of our results. Consider a waveguide formed by $p$-$n$-$p$ (PNP) or superconductor/metal/superconductor (SNS) junction ~\cite{Beenakker2009,1Yliu2018}, the electrons' motion  is  confined inside the middle N region. An electron moving in the waveguide with finite incident angle  would undergo repeated  reflections between the two interfaces of the junction. Due to the presence of the anomalous shift, the dispersion of the confined waveguide mode will be strongly modified ~\cite{Beenakker2009,1Yliu2018}. And our work would be important for  precise description of the dispersion of the waveguide modes of all the momentum points.

\section{Acknowledgement}
The authors thank J. Xun for helpful discussions.
This work was supported by the National Key R\&D Program of China (Grant No. 2020YFA0308800), the NSF of China (Grants Nos. 12234003, 12061131002 and 12004035), and the  National Natural Science Fund for Excellent Young Scientists Fund Program (Overseas) and Beijing Institute of Technology Research Fund Program for Young Scholars.

\bibliography{cite}

\begin{thebibliography}{40}%
\makeatletter
\providecommand \@ifxundefined [1]{%
 \@ifx{#1\undefined}
}%
\providecommand \@ifnum [1]{%
 \ifnum #1\expandafter \@firstoftwo
 \else \expandafter \@secondoftwo
 \fi
}%
\providecommand \@ifx [1]{%
 \ifx #1\expandafter \@firstoftwo
 \else \expandafter \@secondoftwo
 \fi
}%
\providecommand \natexlab [1]{#1}%
\providecommand \enquote  [1]{``#1''}%
\providecommand \bibnamefont  [1]{#1}%
\providecommand \bibfnamefont [1]{#1}%
\providecommand \citenamefont [1]{#1}%
\providecommand \href@noop [0]{\@secondoftwo}%
\providecommand \href [0]{\begingroup \@sanitize@url \@href}%
\providecommand \@href[1]{\@@startlink{#1}\@@href}%
\providecommand \@@href[1]{\endgroup#1\@@endlink}%
\providecommand \@sanitize@url [0]{\catcode `\\12\catcode `\$12\catcode
  `\&12\catcode `\#12\catcode `\^12\catcode `\_12\catcode `\%12\relax}%
\providecommand \@@startlink[1]{}%
\providecommand \@@endlink[0]{}%
\providecommand \url  [0]{\begingroup\@sanitize@url \@url }%
\providecommand \@url [1]{\endgroup\@href {#1}{\urlprefix }}%
\providecommand \urlprefix  [0]{URL }%
\providecommand \Eprint [0]{\href }%
\providecommand \doibase [0]{http://dx.doi.org/}%
\providecommand \selectlanguage [0]{\@gobble}%
\providecommand \bibinfo  [0]{\@secondoftwo}%
\providecommand \bibfield  [0]{\@secondoftwo}%
\providecommand \translation [1]{[#1]}%
\providecommand \BibitemOpen [0]{}%
\providecommand \bibitemStop [0]{}%
\providecommand \bibitemNoStop [0]{.\EOS\space}%
\providecommand \EOS [0]{\spacefactor3000\relax}%
\providecommand \BibitemShut  [1]{\csname bibitem#1\endcsname}%
\let\auto@bib@innerbib\@empty
%</preamble>
\bibitem [{\citenamefont {Bliokh}\ and\ \citenamefont
  {Aiello}(2013)}]{Bliokh_2013}%
  \BibitemOpen
  \bibfield  {author} {\bibinfo {author} {\bibfnamefont {K.~Y.}\ \bibnamefont
  {Bliokh}}\ and\ \bibinfo {author} {\bibfnamefont {A.}~\bibnamefont
  {Aiello}},\ }\href {\doibase 10.1088/2040-8978/15/1/014001} {\bibfield
  {journal} {\bibinfo  {journal} {Journal of Optics}\ }\textbf {\bibinfo
  {volume} {15}},\ \bibinfo {pages} {014001} (\bibinfo {year}
  {2013})}\BibitemShut {NoStop}%
\bibitem [{\citenamefont {{Goos}}\ and\ \citenamefont
  {{H{\"a}nchen}}(1947)}]{Goos1947}%
  \BibitemOpen
  \bibfield  {author} {\bibinfo {author} {\bibfnamefont {F.}~\bibnamefont
  {{Goos}}}\ and\ \bibinfo {author} {\bibfnamefont {H.}~\bibnamefont
  {{H{\"a}nchen}}},\ }\href {\doibase 10.1002/andp.19474360704} {\bibfield
  {journal} {\bibinfo  {journal} {Annalen der Physik}\ }\textbf {\bibinfo
  {volume} {436}},\ \bibinfo {pages} {333} (\bibinfo {year}
  {1947})}\BibitemShut {NoStop}%
\bibitem [{\citenamefont {Renard}(1964)}]{Renard1964}%
  \BibitemOpen
  \bibfield  {author} {\bibinfo {author} {\bibfnamefont {R.~H.}\ \bibnamefont
  {Renard}},\ }\href {\doibase 10.1364/JOSA.54.001190} {\bibfield  {journal}
  {\bibinfo  {journal} {J. Opt. Soc. Am.}\ }\textbf {\bibinfo {volume} {54}},\
  \bibinfo {pages} {1190} (\bibinfo {year} {1964})}\BibitemShut {NoStop}%
\bibitem [{\citenamefont {F.~I.~Fedorov}(1955)}]{IF1955}%
  \BibitemOpen
  \bibfield  {author} {\bibinfo {author} {\bibfnamefont {D.~A.}\ \bibnamefont
  {F.~I.~Fedorov}},\ }\href@noop {} {\bibfield  {journal} {\bibinfo  {journal}
  {Nauk SSSR}\ }\textbf {\bibinfo {volume} {105}},\ \bibinfo {pages} {465}
  (\bibinfo {year} {1955})}\BibitemShut {NoStop}%
\bibitem [{\citenamefont {Imbert}(1972)}]{IF1972}%
  \BibitemOpen
  \bibfield  {author} {\bibinfo {author} {\bibfnamefont {C.}~\bibnamefont
  {Imbert}},\ }\href {\doibase 10.1103/PhysRevD.5.787} {\bibfield  {journal}
  {\bibinfo  {journal} {Phys. Rev. D}\ }\textbf {\bibinfo {volume} {5}},\
  \bibinfo {pages} {787} (\bibinfo {year} {1972})}\BibitemShut {NoStop}%
\bibitem [{\citenamefont {Onoda}\ \emph {et~al.}(2004)\citenamefont {Onoda},
  \citenamefont {Murakami},\ and\ \citenamefont {Nagaosa}}]{Onoda2004}%
  \BibitemOpen
  \bibfield  {author} {\bibinfo {author} {\bibfnamefont {S.}~\bibnamefont
  {Onoda}}, \bibinfo {author} {\bibfnamefont {S.}~\bibnamefont {Murakami}}, \
  and\ \bibinfo {author} {\bibfnamefont {N.}~\bibnamefont {Nagaosa}},\ }\href
  {\doibase 10.1103/PhysRevLett.93.167602} {\bibfield  {journal} {\bibinfo
  {journal} {Phys. Rev. Lett.}\ }\textbf {\bibinfo {volume} {93}},\ \bibinfo
  {pages} {167602} (\bibinfo {year} {2004})}\BibitemShut {NoStop}%
\bibitem [{\citenamefont {Miller}\ and\ \citenamefont
  {Ashby}(1972)}]{Miller1972}%
  \BibitemOpen
  \bibfield  {author} {\bibinfo {author} {\bibfnamefont {S.~C.}\ \bibnamefont
  {Miller}}\ and\ \bibinfo {author} {\bibfnamefont {N.}~\bibnamefont {Ashby}},\
  }\href {\doibase 10.1103/PhysRevLett.29.740} {\bibfield  {journal} {\bibinfo
  {journal} {Phys. Rev. Lett.}\ }\textbf {\bibinfo {volume} {29}},\ \bibinfo
  {pages} {740} (\bibinfo {year} {1972})}\BibitemShut {NoStop}%
\bibitem [{\citenamefont {Beenakker}\ \emph {et~al.}(2009)\citenamefont
  {Beenakker}, \citenamefont {Sepkhanov}, \citenamefont {Akhmerov},\ and\
  \citenamefont {Tworzyd\l{}o}}]{Beenakker2009}%
  \BibitemOpen
  \bibfield  {author} {\bibinfo {author} {\bibfnamefont {C.~W.~J.}\
  \bibnamefont {Beenakker}}, \bibinfo {author} {\bibfnamefont {R.~A.}\
  \bibnamefont {Sepkhanov}}, \bibinfo {author} {\bibfnamefont {A.~R.}\
  \bibnamefont {Akhmerov}}, \ and\ \bibinfo {author} {\bibfnamefont
  {J.}~\bibnamefont {Tworzyd\l{}o}},\ }\href {\doibase
  10.1103/PhysRevLett.102.146804} {\bibfield  {journal} {\bibinfo  {journal}
  {Phys. Rev. Lett.}\ }\textbf {\bibinfo {volume} {102}},\ \bibinfo {pages}
  {146804} (\bibinfo {year} {2009})}\BibitemShut {NoStop}%
\bibitem [{\citenamefont {Chen}\ \emph {et~al.}(2013)\citenamefont {Chen},
  \citenamefont {Lu}, \citenamefont {Ban},\ and\ \citenamefont
  {Li}}]{Chen2013}%
  \BibitemOpen
  \bibfield  {author} {\bibinfo {author} {\bibfnamefont {X.}~\bibnamefont
  {Chen}}, \bibinfo {author} {\bibfnamefont {X.-J.}\ \bibnamefont {Lu}},
  \bibinfo {author} {\bibfnamefont {Y.}~\bibnamefont {Ban}}, \ and\ \bibinfo
  {author} {\bibfnamefont {C.-F.}\ \bibnamefont {Li}},\ }\href@noop {}
  {\bibfield  {journal} {\bibinfo  {journal} {Journal of Optics}\ }\textbf
  {\bibinfo {volume} {15}},\ \bibinfo {pages} {033001} (\bibinfo {year}
  {2013})}\BibitemShut {NoStop}%
\bibitem [{\citenamefont {Wu}\ \emph {et~al.}(2011)\citenamefont {Wu},
  \citenamefont {Zhai}, \citenamefont {Peeters}, \citenamefont {Xu},\ and\
  \citenamefont {Chang}}]{Wu2011}%
  \BibitemOpen
  \bibfield  {author} {\bibinfo {author} {\bibfnamefont {Z.}~\bibnamefont
  {Wu}}, \bibinfo {author} {\bibfnamefont {F.}~\bibnamefont {Zhai}}, \bibinfo
  {author} {\bibfnamefont {F.~M.}\ \bibnamefont {Peeters}}, \bibinfo {author}
  {\bibfnamefont {H.~Q.}\ \bibnamefont {Xu}}, \ and\ \bibinfo {author}
  {\bibfnamefont {K.}~\bibnamefont {Chang}},\ }\href {\doibase
  10.1103/PhysRevLett.106.176802} {\bibfield  {journal} {\bibinfo  {journal}
  {Phys. Rev. Lett.}\ }\textbf {\bibinfo {volume} {106}},\ \bibinfo {pages}
  {176802} (\bibinfo {year} {2011})}\BibitemShut {NoStop}%
\bibitem [{\citenamefont {Yu}\ \emph {et~al.}(2019)\citenamefont {Yu},
  \citenamefont {Liu},\ and\ \citenamefont {Yang}}]{Yu2019}%
  \BibitemOpen
  \bibfield  {author} {\bibinfo {author} {\bibfnamefont {Z.-M.}\ \bibnamefont
  {Yu}}, \bibinfo {author} {\bibfnamefont {Y.}~\bibnamefont {Liu}}, \ and\
  \bibinfo {author} {\bibfnamefont {S.~A.}\ \bibnamefont {Yang}},\ }\href
  {\doibase 10.1007/s11467-019-0882-7} {\bibfield  {journal} {\bibinfo
  {journal} {Frontiers of Physics}\ }\textbf {\bibinfo {volume} {14}},\
  \bibinfo {pages} {33402} (\bibinfo {year} {2019})}\BibitemShut {NoStop}%
\bibitem [{\citenamefont {Huang}\ \emph {et~al.}(2008)\citenamefont {Huang},
  \citenamefont {Duan}, \citenamefont {Ling},\ and\ \citenamefont
  {Zhang}}]{JH2008}%
  \BibitemOpen
  \bibfield  {author} {\bibinfo {author} {\bibfnamefont {J.}~\bibnamefont
  {Huang}}, \bibinfo {author} {\bibfnamefont {Z.}~\bibnamefont {Duan}},
  \bibinfo {author} {\bibfnamefont {H.~Y.}\ \bibnamefont {Ling}}, \ and\
  \bibinfo {author} {\bibfnamefont {W.}~\bibnamefont {Zhang}},\ }\href
  {\doibase 10.1103/PhysRevA.77.063608} {\bibfield  {journal} {\bibinfo
  {journal} {Phys. Rev. A}\ }\textbf {\bibinfo {volume} {77}},\ \bibinfo
  {pages} {063608} (\bibinfo {year} {2008})}\BibitemShut {NoStop}%
\bibitem [{\citenamefont {de~Haan}\ \emph {et~al.}(2010)\citenamefont
  {de~Haan}, \citenamefont {Plomp}, \citenamefont {Rekveldt}, \citenamefont
  {Kraan}, \citenamefont {van Well}, \citenamefont {Dalgliesh},\ and\
  \citenamefont {Langridge}}]{Haan2010}%
  \BibitemOpen
  \bibfield  {author} {\bibinfo {author} {\bibfnamefont {V.-O.}\ \bibnamefont
  {de~Haan}}, \bibinfo {author} {\bibfnamefont {J.}~\bibnamefont {Plomp}},
  \bibinfo {author} {\bibfnamefont {T.~M.}\ \bibnamefont {Rekveldt}}, \bibinfo
  {author} {\bibfnamefont {W.~H.}\ \bibnamefont {Kraan}}, \bibinfo {author}
  {\bibfnamefont {A.~A.}\ \bibnamefont {van Well}}, \bibinfo {author}
  {\bibfnamefont {R.~M.}\ \bibnamefont {Dalgliesh}}, \ and\ \bibinfo {author}
  {\bibfnamefont {S.}~\bibnamefont {Langridge}},\ }\href {\doibase
  10.1103/PhysRevLett.104.010401} {\bibfield  {journal} {\bibinfo  {journal}
  {Phys. Rev. Lett.}\ }\textbf {\bibinfo {volume} {104}},\ \bibinfo {pages}
  {010401} (\bibinfo {year} {2010})}\BibitemShut {NoStop}%
\bibitem [{\citenamefont {Chiu}\ \emph {et~al.}(2016)\citenamefont {Chiu},
  \citenamefont {Teo}, \citenamefont {Schnyder},\ and\ \citenamefont
  {Ryu}}]{Chiu2016}%
  \BibitemOpen
  \bibfield  {author} {\bibinfo {author} {\bibfnamefont {C.-K.}\ \bibnamefont
  {Chiu}}, \bibinfo {author} {\bibfnamefont {J.~C.~Y.}\ \bibnamefont {Teo}},
  \bibinfo {author} {\bibfnamefont {A.~P.}\ \bibnamefont {Schnyder}}, \ and\
  \bibinfo {author} {\bibfnamefont {S.}~\bibnamefont {Ryu}},\ }\href {\doibase
  10.1103/RevModPhys.88.035005} {\bibfield  {journal} {\bibinfo  {journal}
  {Rev. Mod. Phys.}\ }\textbf {\bibinfo {volume} {88}},\ \bibinfo {pages}
  {035005} (\bibinfo {year} {2016})}\BibitemShut {NoStop}%
\bibitem [{\citenamefont {Armitage}\ \emph {et~al.}(2018)\citenamefont
  {Armitage}, \citenamefont {Mele},\ and\ \citenamefont
  {Vishwanath}}]{Armitage2018}%
  \BibitemOpen
  \bibfield  {author} {\bibinfo {author} {\bibfnamefont {N.~P.}\ \bibnamefont
  {Armitage}}, \bibinfo {author} {\bibfnamefont {E.~J.}\ \bibnamefont {Mele}},
  \ and\ \bibinfo {author} {\bibfnamefont {A.}~\bibnamefont {Vishwanath}},\
  }\href {\doibase 10.1103/RevModPhys.90.015001} {\bibfield  {journal}
  {\bibinfo  {journal} {Rev. Mod. Phys.}\ }\textbf {\bibinfo {volume} {90}},\
  \bibinfo {pages} {015001} (\bibinfo {year} {2018})}\BibitemShut {NoStop}%
\bibitem [{\citenamefont {Wan}\ \emph {et~al.}(2011)\citenamefont {Wan},
  \citenamefont {Turner}, \citenamefont {Vishwanath},\ and\ \citenamefont
  {Savrasov}}]{Wan2011}%
  \BibitemOpen
  \bibfield  {author} {\bibinfo {author} {\bibfnamefont {X.}~\bibnamefont
  {Wan}}, \bibinfo {author} {\bibfnamefont {A.~M.}\ \bibnamefont {Turner}},
  \bibinfo {author} {\bibfnamefont {A.}~\bibnamefont {Vishwanath}}, \ and\
  \bibinfo {author} {\bibfnamefont {S.~Y.}\ \bibnamefont {Savrasov}},\ }\href
  {\doibase 10.1103/PhysRevB.83.205101} {\bibfield  {journal} {\bibinfo
  {journal} {Phys. Rev. B}\ }\textbf {\bibinfo {volume} {83}},\ \bibinfo
  {pages} {205101} (\bibinfo {year} {2011})}\BibitemShut {NoStop}%
\bibitem [{\citenamefont {Yu}\ \emph {et~al.}(2016)\citenamefont {Yu},
  \citenamefont {Yao},\ and\ \citenamefont {Yang}}]{yu2016}%
  \BibitemOpen
  \bibfield  {author} {\bibinfo {author} {\bibfnamefont {Z.-M.}\ \bibnamefont
  {Yu}}, \bibinfo {author} {\bibfnamefont {Y.}~\bibnamefont {Yao}}, \ and\
  \bibinfo {author} {\bibfnamefont {S.~A.}\ \bibnamefont {Yang}},\ }\href
  {\doibase 10.1103/PhysRevLett.117.077202} {\bibfield  {journal} {\bibinfo
  {journal} {Phys. Rev. Lett.}\ }\textbf {\bibinfo {volume} {117}},\ \bibinfo
  {pages} {077202} (\bibinfo {year} {2016})}\BibitemShut {NoStop}%
\bibitem [{\citenamefont {Weng}\ \emph {et~al.}(2015)\citenamefont {Weng},
  \citenamefont {Liang}, \citenamefont {Xu}, \citenamefont {Yu}, \citenamefont
  {Fang}, \citenamefont {Dai},\ and\ \citenamefont {Kawazoe}}]{Weng2015}%
  \BibitemOpen
  \bibfield  {author} {\bibinfo {author} {\bibfnamefont {H.}~\bibnamefont
  {Weng}}, \bibinfo {author} {\bibfnamefont {Y.}~\bibnamefont {Liang}},
  \bibinfo {author} {\bibfnamefont {Q.}~\bibnamefont {Xu}}, \bibinfo {author}
  {\bibfnamefont {R.}~\bibnamefont {Yu}}, \bibinfo {author} {\bibfnamefont
  {Z.}~\bibnamefont {Fang}}, \bibinfo {author} {\bibfnamefont {X.}~\bibnamefont
  {Dai}}, \ and\ \bibinfo {author} {\bibfnamefont {Y.}~\bibnamefont
  {Kawazoe}},\ }\href {\doibase 10.1103/PhysRevB.92.045108} {\bibfield
  {journal} {\bibinfo  {journal} {Phys. Rev. B}\ }\textbf {\bibinfo {volume}
  {92}},\ \bibinfo {pages} {045108} (\bibinfo {year} {2015})}\BibitemShut
  {NoStop}%
\bibitem [{\citenamefont {Wu}\ \emph {et~al.}(2018)\citenamefont {Wu},
  \citenamefont {Liu}, \citenamefont {Li}, \citenamefont {Zhong}, \citenamefont
  {Yu}, \citenamefont {Sheng}, \citenamefont {Zhao},\ and\ \citenamefont
  {Yang}}]{Wu2018}%
  \BibitemOpen
  \bibfield  {author} {\bibinfo {author} {\bibfnamefont {W.}~\bibnamefont
  {Wu}}, \bibinfo {author} {\bibfnamefont {Y.}~\bibnamefont {Liu}}, \bibinfo
  {author} {\bibfnamefont {S.}~\bibnamefont {Li}}, \bibinfo {author}
  {\bibfnamefont {C.}~\bibnamefont {Zhong}}, \bibinfo {author} {\bibfnamefont
  {Z.-M.}\ \bibnamefont {Yu}}, \bibinfo {author} {\bibfnamefont {X.-L.}\
  \bibnamefont {Sheng}}, \bibinfo {author} {\bibfnamefont {Y.~X.}\ \bibnamefont
  {Zhao}}, \ and\ \bibinfo {author} {\bibfnamefont {S.~A.}\ \bibnamefont
  {Yang}},\ }\href {\doibase 10.1103/PhysRevB.97.115125} {\bibfield  {journal}
  {\bibinfo  {journal} {Phys. Rev. B}\ }\textbf {\bibinfo {volume} {97}},\
  \bibinfo {pages} {115125} (\bibinfo {year} {2018})}\BibitemShut {NoStop}%
\bibitem [{\citenamefont {Soluyanov}\ \emph {et~al.}(2015)\citenamefont
  {Soluyanov}, \citenamefont {Gresch}, \citenamefont {Wang}, \citenamefont
  {Wu}, \citenamefont {Troyer}, \citenamefont {Dai},\ and\ \citenamefont
  {Bernevig}}]{Soluyanov2015}%
  \BibitemOpen
  \bibfield  {author} {\bibinfo {author} {\bibfnamefont {A.~A.}\ \bibnamefont
  {Soluyanov}}, \bibinfo {author} {\bibfnamefont {D.}~\bibnamefont {Gresch}},
  \bibinfo {author} {\bibfnamefont {Z.}~\bibnamefont {Wang}}, \bibinfo {author}
  {\bibfnamefont {Q.}~\bibnamefont {Wu}}, \bibinfo {author} {\bibfnamefont
  {M.}~\bibnamefont {Troyer}}, \bibinfo {author} {\bibfnamefont
  {X.}~\bibnamefont {Dai}}, \ and\ \bibinfo {author} {\bibfnamefont {B.~A.}\
  \bibnamefont {Bernevig}},\ }\href@noop {} {\bibfield  {journal} {\bibinfo
  {journal} {Nature}\ }\textbf {\bibinfo {volume} {527}},\ \bibinfo {pages}
  {495} (\bibinfo {year} {2015})}\BibitemShut {NoStop}%
\bibitem [{\citenamefont {Li}\ \emph {et~al.}(2021{\natexlab{a}})\citenamefont
  {Li}, \citenamefont {Deng}, \citenamefont {Fu}, \citenamefont {Li},
  \citenamefont {Ma}, \citenamefont {Han}, \citenamefont {Zhou}, \citenamefont
  {Zhou},\ and\ \citenamefont {Yao}}]{Lix2021}%
  \BibitemOpen
  \bibfield  {author} {\bibinfo {author} {\bibfnamefont {X.-P.}\ \bibnamefont
  {Li}}, \bibinfo {author} {\bibfnamefont {K.}~\bibnamefont {Deng}}, \bibinfo
  {author} {\bibfnamefont {B.}~\bibnamefont {Fu}}, \bibinfo {author}
  {\bibfnamefont {Y.}~\bibnamefont {Li}}, \bibinfo {author} {\bibfnamefont
  {D.-S.}\ \bibnamefont {Ma}}, \bibinfo {author} {\bibfnamefont
  {J.}~\bibnamefont {Han}}, \bibinfo {author} {\bibfnamefont {J.}~\bibnamefont
  {Zhou}}, \bibinfo {author} {\bibfnamefont {S.}~\bibnamefont {Zhou}}, \ and\
  \bibinfo {author} {\bibfnamefont {Y.}~\bibnamefont {Yao}},\ }\href {\doibase
  10.1103/PhysRevB.103.L081402} {\bibfield  {journal} {\bibinfo  {journal}
  {Phys. Rev. B}\ }\textbf {\bibinfo {volume} {103}},\ \bibinfo {pages}
  {L081402} (\bibinfo {year} {2021}{\natexlab{a}})}\BibitemShut {NoStop}%
\bibitem [{\citenamefont {Li}\ \emph {et~al.}(2021{\natexlab{b}})\citenamefont
  {Li}, \citenamefont {Fu}, \citenamefont {Ma}, \citenamefont {Cui},
  \citenamefont {Yu},\ and\ \citenamefont {Yao}}]{2Lix2021}%
  \BibitemOpen
  \bibfield  {author} {\bibinfo {author} {\bibfnamefont {X.-P.}\ \bibnamefont
  {Li}}, \bibinfo {author} {\bibfnamefont {B.}~\bibnamefont {Fu}}, \bibinfo
  {author} {\bibfnamefont {D.-S.}\ \bibnamefont {Ma}}, \bibinfo {author}
  {\bibfnamefont {C.}~\bibnamefont {Cui}}, \bibinfo {author} {\bibfnamefont
  {Z.-M.}\ \bibnamefont {Yu}}, \ and\ \bibinfo {author} {\bibfnamefont
  {Y.}~\bibnamefont {Yao}},\ }\href {\doibase 10.1103/PhysRevB.103.L161109}
  {\bibfield  {journal} {\bibinfo  {journal} {Phys. Rev. B}\ }\textbf {\bibinfo
  {volume} {103}},\ \bibinfo {pages} {L161109} (\bibinfo {year}
  {2021}{\natexlab{b}})}\BibitemShut {NoStop}%
\bibitem [{\citenamefont {Yu}\ \emph {et~al.}(2022)\citenamefont {Yu},
  \citenamefont {Zhang}, \citenamefont {Liu}, \citenamefont {Wu}, \citenamefont
  {Li}, \citenamefont {Zhang}, \citenamefont {Yang},\ and\ \citenamefont
  {Yao}}]{yu2022}%
  \BibitemOpen
  \bibfield  {author} {\bibinfo {author} {\bibfnamefont {Z.-M.}\ \bibnamefont
  {Yu}}, \bibinfo {author} {\bibfnamefont {Z.}~\bibnamefont {Zhang}}, \bibinfo
  {author} {\bibfnamefont {G.-B.}\ \bibnamefont {Liu}}, \bibinfo {author}
  {\bibfnamefont {W.}~\bibnamefont {Wu}}, \bibinfo {author} {\bibfnamefont
  {X.-P.}\ \bibnamefont {Li}}, \bibinfo {author} {\bibfnamefont {R.-W.}\
  \bibnamefont {Zhang}}, \bibinfo {author} {\bibfnamefont {S.~A.}\ \bibnamefont
  {Yang}}, \ and\ \bibinfo {author} {\bibfnamefont {Y.}~\bibnamefont {Yao}},\
  }\href@noop {} {\bibfield  {journal} {\bibinfo  {journal} {Science Bulletin}\
  }\textbf {\bibinfo {volume} {67}},\ \bibinfo {pages} {375} (\bibinfo {year}
  {2022})}\BibitemShut {NoStop}%
\bibitem [{\citenamefont {Liu}\ \emph {et~al.}(2022)\citenamefont {Liu},
  \citenamefont {Zhang}, \citenamefont {Yu}, \citenamefont {Yang},\ and\
  \citenamefont {Yao}}]{Guibin2022}%
  \BibitemOpen
  \bibfield  {author} {\bibinfo {author} {\bibfnamefont {G.-B.}\ \bibnamefont
  {Liu}}, \bibinfo {author} {\bibfnamefont {Z.}~\bibnamefont {Zhang}}, \bibinfo
  {author} {\bibfnamefont {Z.-M.}\ \bibnamefont {Yu}}, \bibinfo {author}
  {\bibfnamefont {S.~A.}\ \bibnamefont {Yang}}, \ and\ \bibinfo {author}
  {\bibfnamefont {Y.}~\bibnamefont {Yao}},\ }\href {\doibase
  10.1103/PhysRevB.105.085117} {\bibfield  {journal} {\bibinfo  {journal}
  {Phys. Rev. B}\ }\textbf {\bibinfo {volume} {105}},\ \bibinfo {pages}
  {085117} (\bibinfo {year} {2022})}\BibitemShut {NoStop}%
\bibitem [{\citenamefont {Zhang}\ \emph {et~al.}(2022)\citenamefont {Zhang},
  \citenamefont {Liu}, \citenamefont {Yu}, \citenamefont {Yang},\ and\
  \citenamefont {Yao}}]{Zeying2022}%
  \BibitemOpen
  \bibfield  {author} {\bibinfo {author} {\bibfnamefont {Z.}~\bibnamefont
  {Zhang}}, \bibinfo {author} {\bibfnamefont {G.-B.}\ \bibnamefont {Liu}},
  \bibinfo {author} {\bibfnamefont {Z.-M.}\ \bibnamefont {Yu}}, \bibinfo
  {author} {\bibfnamefont {S.~A.}\ \bibnamefont {Yang}}, \ and\ \bibinfo
  {author} {\bibfnamefont {Y.}~\bibnamefont {Yao}},\ }\href {\doibase
  10.1103/PhysRevB.105.104426} {\bibfield  {journal} {\bibinfo  {journal}
  {Phys. Rev. B}\ }\textbf {\bibinfo {volume} {105}},\ \bibinfo {pages}
  {104426} (\bibinfo {year} {2022})}\BibitemShut {NoStop}%
\bibitem [{\citenamefont {Jiang}\ \emph {et~al.}(2015)\citenamefont {Jiang},
  \citenamefont {Jiang}, \citenamefont {Liu}, \citenamefont {Sun},\ and\
  \citenamefont {Xie}}]{Jiang2015}%
  \BibitemOpen
  \bibfield  {author} {\bibinfo {author} {\bibfnamefont {Q.-D.}\ \bibnamefont
  {Jiang}}, \bibinfo {author} {\bibfnamefont {H.}~\bibnamefont {Jiang}},
  \bibinfo {author} {\bibfnamefont {H.}~\bibnamefont {Liu}}, \bibinfo {author}
  {\bibfnamefont {Q.-F.}\ \bibnamefont {Sun}}, \ and\ \bibinfo {author}
  {\bibfnamefont {X.~C.}\ \bibnamefont {Xie}},\ }\href {\doibase
  10.1103/PhysRevLett.115.156602} {\bibfield  {journal} {\bibinfo  {journal}
  {Phys. Rev. Lett.}\ }\textbf {\bibinfo {volume} {115}},\ \bibinfo {pages}
  {156602} (\bibinfo {year} {2015})}\BibitemShut {NoStop}%
\bibitem [{\citenamefont {Yang}\ \emph {et~al.}(2015)\citenamefont {Yang},
  \citenamefont {Pan},\ and\ \citenamefont {Zhang}}]{Yang2015}%
  \BibitemOpen
  \bibfield  {author} {\bibinfo {author} {\bibfnamefont {S.~A.}\ \bibnamefont
  {Yang}}, \bibinfo {author} {\bibfnamefont {H.}~\bibnamefont {Pan}}, \ and\
  \bibinfo {author} {\bibfnamefont {F.}~\bibnamefont {Zhang}},\ }\href
  {\doibase 10.1103/PhysRevLett.115.156603} {\bibfield  {journal} {\bibinfo
  {journal} {Phys. Rev. Lett.}\ }\textbf {\bibinfo {volume} {115}},\ \bibinfo
  {pages} {156603} (\bibinfo {year} {2015})}\BibitemShut {NoStop}%
\bibitem [{\citenamefont {Hao}\ \emph {et~al.}(2019)\citenamefont {Hao},
  \citenamefont {Wang},\ and\ \citenamefont {Yao}}]{Yao2019}%
  \BibitemOpen
  \bibfield  {author} {\bibinfo {author} {\bibfnamefont {Y.-R.}\ \bibnamefont
  {Hao}}, \bibinfo {author} {\bibfnamefont {L.}~\bibnamefont {Wang}}, \ and\
  \bibinfo {author} {\bibfnamefont {D.-X.}\ \bibnamefont {Yao}},\ }\href
  {\doibase 10.1103/PhysRevB.99.165406} {\bibfield  {journal} {\bibinfo
  {journal} {Phys. Rev. B}\ }\textbf {\bibinfo {volume} {99}},\ \bibinfo
  {pages} {165406} (\bibinfo {year} {2019})}\BibitemShut {NoStop}%
\bibitem [{\citenamefont {Feng}\ \emph {et~al.}(2020)\citenamefont {Feng},
  \citenamefont {Liu}, \citenamefont {Yu}, \citenamefont {Ma}, \citenamefont
  {Ang}, \citenamefont {Ang},\ and\ \citenamefont {Yang}}]{Feng2020}%
  \BibitemOpen
  \bibfield  {author} {\bibinfo {author} {\bibfnamefont {X.}~\bibnamefont
  {Feng}}, \bibinfo {author} {\bibfnamefont {Y.}~\bibnamefont {Liu}}, \bibinfo
  {author} {\bibfnamefont {Z.-M.}\ \bibnamefont {Yu}}, \bibinfo {author}
  {\bibfnamefont {Z.}~\bibnamefont {Ma}}, \bibinfo {author} {\bibfnamefont
  {L.~K.}\ \bibnamefont {Ang}}, \bibinfo {author} {\bibfnamefont {Y.~S.}\
  \bibnamefont {Ang}}, \ and\ \bibinfo {author} {\bibfnamefont {S.~A.}\
  \bibnamefont {Yang}},\ }\href {\doibase 10.1103/PhysRevB.101.235417}
  {\bibfield  {journal} {\bibinfo  {journal} {Phys. Rev. B}\ }\textbf {\bibinfo
  {volume} {101}},\ \bibinfo {pages} {235417} (\bibinfo {year}
  {2020})}\BibitemShut {NoStop}%
\bibitem [{\citenamefont {Liu}\ \emph {et~al.}(2020)\citenamefont {Liu},
  \citenamefont {Yu}, \citenamefont {Xiao},\ and\ \citenamefont
  {Yang}}]{Yliu2020}%
  \BibitemOpen
  \bibfield  {author} {\bibinfo {author} {\bibfnamefont {Y.}~\bibnamefont
  {Liu}}, \bibinfo {author} {\bibfnamefont {Z.-M.}\ \bibnamefont {Yu}},
  \bibinfo {author} {\bibfnamefont {C.}~\bibnamefont {Xiao}}, \ and\ \bibinfo
  {author} {\bibfnamefont {S.~A.}\ \bibnamefont {Yang}},\ }\href {\doibase
  10.1103/PhysRevLett.125.076801} {\bibfield  {journal} {\bibinfo  {journal}
  {Phys. Rev. Lett.}\ }\textbf {\bibinfo {volume} {125}},\ \bibinfo {pages}
  {076801} (\bibinfo {year} {2020})}\BibitemShut {NoStop}%
\bibitem [{\citenamefont {Liu}\ \emph {et~al.}()\citenamefont {Liu},
  \citenamefont {Yu},\ and\ \citenamefont {Yang}}]{Yliu2017}%
  \BibitemOpen
  \bibfield  {author} {\bibinfo {author} {\bibfnamefont {Y.}~\bibnamefont
  {Liu}}, \bibinfo {author} {\bibfnamefont {Z.-M.}\ \bibnamefont {Yu}}, \ and\
  \bibinfo {author} {\bibfnamefont {S.~A.}\ \bibnamefont {Yang}},\ }\href@noop
  {} {\bibfield  {journal} {\bibinfo  {journal} {Phys. Rev. B}\ }\textbf
  {\bibinfo {volume} {96}},\ \bibinfo {pages} {121101}}\BibitemShut {NoStop}%
\bibitem [{\citenamefont {Liu}\ \emph {et~al.}(2018{\natexlab{a}})\citenamefont
  {Liu}, \citenamefont {Yu}, \citenamefont {Jiang},\ and\ \citenamefont
  {Yang}}]{1Yliu2018}%
  \BibitemOpen
  \bibfield  {author} {\bibinfo {author} {\bibfnamefont {Y.}~\bibnamefont
  {Liu}}, \bibinfo {author} {\bibfnamefont {Z.-M.}\ \bibnamefont {Yu}},
  \bibinfo {author} {\bibfnamefont {H.}~\bibnamefont {Jiang}}, \ and\ \bibinfo
  {author} {\bibfnamefont {S.~A.}\ \bibnamefont {Yang}},\ }\href {\doibase
  10.1103/PhysRevB.98.075151} {\bibfield  {journal} {\bibinfo  {journal} {Phys.
  Rev. B}\ }\textbf {\bibinfo {volume} {98}},\ \bibinfo {pages} {075151}
  (\bibinfo {year} {2018}{\natexlab{a}})}\BibitemShut {NoStop}%
\bibitem [{\citenamefont {Yu}\ \emph {et~al.}(2018)\citenamefont {Yu},
  \citenamefont {Liu}, \citenamefont {Yao},\ and\ \citenamefont
  {Yang}}]{Yu2018}%
  \BibitemOpen
  \bibfield  {author} {\bibinfo {author} {\bibfnamefont {Z.-M.}\ \bibnamefont
  {Yu}}, \bibinfo {author} {\bibfnamefont {Y.}~\bibnamefont {Liu}}, \bibinfo
  {author} {\bibfnamefont {Y.}~\bibnamefont {Yao}}, \ and\ \bibinfo {author}
  {\bibfnamefont {S.~A.}\ \bibnamefont {Yang}},\ }\href {\doibase
  10.1103/PhysRevLett.121.176602} {\bibfield  {journal} {\bibinfo  {journal}
  {Phys. Rev. Lett.}\ }\textbf {\bibinfo {volume} {121}},\ \bibinfo {pages}
  {176602} (\bibinfo {year} {2018})}\BibitemShut {NoStop}%
\bibitem [{\citenamefont {Liu}\ \emph {et~al.}(2018{\natexlab{b}})\citenamefont
  {Liu}, \citenamefont {Yu}, \citenamefont {Liu}, \citenamefont {Jiang},\ and\
  \citenamefont {Yang}}]{2Yliu2018}%
  \BibitemOpen
  \bibfield  {author} {\bibinfo {author} {\bibfnamefont {Y.}~\bibnamefont
  {Liu}}, \bibinfo {author} {\bibfnamefont {Z.-M.}\ \bibnamefont {Yu}},
  \bibinfo {author} {\bibfnamefont {J.}~\bibnamefont {Liu}}, \bibinfo {author}
  {\bibfnamefont {H.}~\bibnamefont {Jiang}}, \ and\ \bibinfo {author}
  {\bibfnamefont {S.~A.}\ \bibnamefont {Yang}},\ }\href {\doibase
  10.1103/PhysRevB.98.195141} {\bibfield  {journal} {\bibinfo  {journal} {Phys.
  Rev. B}\ }\textbf {\bibinfo {volume} {98}},\ \bibinfo {pages} {195141}
  (\bibinfo {year} {2018}{\natexlab{b}})}\BibitemShut {NoStop}%
\bibitem [{\citenamefont {Luo}\ \emph {et~al.}(2011)\citenamefont {Luo},
  \citenamefont {Zhou}, \citenamefont {Shu}, \citenamefont {Wen},\ and\
  \citenamefont {Fan}}]{Luo2011}%
  \BibitemOpen
  \bibfield  {author} {\bibinfo {author} {\bibfnamefont {H.}~\bibnamefont
  {Luo}}, \bibinfo {author} {\bibfnamefont {X.}~\bibnamefont {Zhou}}, \bibinfo
  {author} {\bibfnamefont {W.}~\bibnamefont {Shu}}, \bibinfo {author}
  {\bibfnamefont {S.}~\bibnamefont {Wen}}, \ and\ \bibinfo {author}
  {\bibfnamefont {D.}~\bibnamefont {Fan}},\ }\href {\doibase
  10.1103/PhysRevA.84.043806} {\bibfield  {journal} {\bibinfo  {journal} {Phys.
  Rev. A}\ }\textbf {\bibinfo {volume} {84}},\ \bibinfo {pages} {043806}
  (\bibinfo {year} {2011})}\BibitemShut {NoStop}%
\bibitem [{\citenamefont {Ling}\ \emph {et~al.}(2021)\citenamefont {Ling},
  \citenamefont {Xiao}, \citenamefont {Chen}, \citenamefont {Zhou},
  \citenamefont {Luo},\ and\ \citenamefont {Zhou}}]{Ling2021}%
  \BibitemOpen
  \bibfield  {author} {\bibinfo {author} {\bibfnamefont {X.}~\bibnamefont
  {Ling}}, \bibinfo {author} {\bibfnamefont {W.}~\bibnamefont {Xiao}}, \bibinfo
  {author} {\bibfnamefont {S.}~\bibnamefont {Chen}}, \bibinfo {author}
  {\bibfnamefont {X.}~\bibnamefont {Zhou}}, \bibinfo {author} {\bibfnamefont
  {H.}~\bibnamefont {Luo}}, \ and\ \bibinfo {author} {\bibfnamefont
  {L.}~\bibnamefont {Zhou}},\ }\href {\doibase 10.1103/PhysRevA.103.033515}
  {\bibfield  {journal} {\bibinfo  {journal} {Phys. Rev. A}\ }\textbf {\bibinfo
  {volume} {103}},\ \bibinfo {pages} {033515} (\bibinfo {year}
  {2021})}\BibitemShut {NoStop}%
\bibitem [{\citenamefont {Yu}\ \emph {et~al.}(2017)\citenamefont {Yu},
  \citenamefont {Ma}, \citenamefont {Pan},\ and\ \citenamefont {Yao}}]{Ma2018}%
  \BibitemOpen
  \bibfield  {author} {\bibinfo {author} {\bibfnamefont {Z.-M.}\ \bibnamefont
  {Yu}}, \bibinfo {author} {\bibfnamefont {D.-S.}\ \bibnamefont {Ma}}, \bibinfo
  {author} {\bibfnamefont {H.}~\bibnamefont {Pan}}, \ and\ \bibinfo {author}
  {\bibfnamefont {Y.}~\bibnamefont {Yao}},\ }\href {\doibase
  10.1103/PhysRevB.96.125152} {\bibfield  {journal} {\bibinfo  {journal} {Phys.
  Rev. B}\ }\textbf {\bibinfo {volume} {96}},\ \bibinfo {pages} {125152}
  (\bibinfo {year} {2017})}\BibitemShut {NoStop}%
\bibitem [{\citenamefont {Lai}\ \emph {et~al.}(1986)\citenamefont {Lai},
  \citenamefont {Cheng},\ and\ \citenamefont {Tang}}]{Lai1986}%
  \BibitemOpen
  \bibfield  {author} {\bibinfo {author} {\bibfnamefont {H.~M.}\ \bibnamefont
  {Lai}}, \bibinfo {author} {\bibfnamefont {F.~C.}\ \bibnamefont {Cheng}}, \
  and\ \bibinfo {author} {\bibfnamefont {W.~K.}\ \bibnamefont {Tang}},\ }\href
  {\doibase 10.1364/JOSAA.3.000550} {\bibfield  {journal} {\bibinfo  {journal}
  {J. Opt. Soc. Am. A}\ }\textbf {\bibinfo {volume} {3}},\ \bibinfo {pages}
  {550} (\bibinfo {year} {1986})}\BibitemShut {NoStop}%
\bibitem [{\citenamefont {Wang}\ \emph {et~al.}(2013)\citenamefont {Wang},
  \citenamefont {Zhu},\ and\ \citenamefont {Zubairy}}]{Wang2013}%
  \BibitemOpen
  \bibfield  {author} {\bibinfo {author} {\bibfnamefont {L.-G.}\ \bibnamefont
  {Wang}}, \bibinfo {author} {\bibfnamefont {S.-Y.}\ \bibnamefont {Zhu}}, \
  and\ \bibinfo {author} {\bibfnamefont {M.~S.}\ \bibnamefont {Zubairy}},\
  }\href {\doibase 10.1103/PhysRevLett.111.223901} {\bibfield  {journal}
  {\bibinfo  {journal} {Phys. Rev. Lett.}\ }\textbf {\bibinfo {volume} {111}},\
  \bibinfo {pages} {223901} (\bibinfo {year} {2013})}\BibitemShut {NoStop}%
\bibitem [{\citenamefont {Qiu}\ \emph {et~al.}(2015)\citenamefont {Qiu},
  \citenamefont {Xie}, \citenamefont {Qiu}, \citenamefont {Zhang},
  \citenamefont {Du},\ and\ \citenamefont {Gao}}]{Qiu2015}%
  \BibitemOpen
  \bibfield  {author} {\bibinfo {author} {\bibfnamefont {X.}~\bibnamefont
  {Qiu}}, \bibinfo {author} {\bibfnamefont {L.}~\bibnamefont {Xie}}, \bibinfo
  {author} {\bibfnamefont {J.}~\bibnamefont {Qiu}}, \bibinfo {author}
  {\bibfnamefont {Z.}~\bibnamefont {Zhang}}, \bibinfo {author} {\bibfnamefont
  {J.}~\bibnamefont {Du}}, \ and\ \bibinfo {author} {\bibfnamefont
  {F.}~\bibnamefont {Gao}},\ }\href {\doibase 10.1364/OE.23.018823} {\bibfield
  {journal} {\bibinfo  {journal} {Opt. Express}\ }\textbf {\bibinfo {volume}
  {23}},\ \bibinfo {pages} {18823} (\bibinfo {year} {2015})}\BibitemShut
  {NoStop}%
\end{thebibliography}%



\end{document}